\documentclass[manuscript]{acmart}
\renewcommand\footnotetextcopyrightpermission[1]{} % removes footnote with conference information in first column
\usepackage{pifont}
\usepackage{amssymb}
\newcommand{\xmark}{\ding{55}}%
\def\BibTeX{{\rm B\kern-.05em{\sc i\kern-.025em b}\kern-.08emT\kern-.1667em\lower.7ex\hbox{E}\kern-.125emX}}
    
\begin{document}

%
% The "title" command has an optional parameter, allowing the author to define a "short title" to be used in page headers.
\title{Hardware and software co-optimization for the initialization failure of the ReRAM based cross-bar array}
\newcommand{\IBMYKT}[0]{\affiliation{IBM Thomas J. Watson Research Center, Yorktown Heights, NY 10598, USA}}
\newcommand{\IBMANT}[0]{\affiliation{IBM Research, Albany, NY 12203, USA}}
\newcommand{\HYU}[0]{\affiliation{Department of Electronic Engineering, Hanyang University, Seoul, South Korea}}
\newcommand{\POSTECH}[0]{\affiliation{Department of Materials Science and Engineering, POSTECH, Pohang, South Korea}}
%
% The "author" command and its associated commands are used to define the authors and their affiliations.
% Of note is the shared affiliation of the first two authors, and the "authornote" and "authornotemark" commands
% used to denote shared contribution to the research.
\author{Youngseok Kim}\IBMANT
\author{Seyoung Kim}\POSTECH \authornote{This work is done when the author is a member of IBM Thomas J. Watson Research Center, Yorktown Heights, NY 10598, USA}
\author{Chun-chen Yeh}\IBMANT\IBMYKT
\author{Vijay Narayanan}\IBMYKT
\author{Jungwook Choi}\HYU \authornote{This work is done when the author is a member of IBM Thomas J. Watson Research Center, Yorktown Heights, NY 10598, USA}
% By default, the full list of authors will be used in the page headers. Often, this list is too long, and will overlap
% other information printed in the page headers. This command allows the author to define a more concise list
% of authors' names for this purpose.

%
% The abstract is a short summary of the work to be presented in the article.
\begin{abstract}
Recent advances in deep neural network demand more than millions of parameters to handle and mandate the high-performance computing resources with improved efficiency. The cross-bar array architecture has been considered as one of the promising deep learning architectures that shows a significant computing gain over the conventional processors. 
To investigate the feasibility of the architecture, we examine non-idealities and their impact on the performance. Specifically, we study the impact of failed cells due to the initialization process of the resistive memory based cross-bar array. Unlike the conventional memory array, individual memory elements cannot be rerouted and, thus, may have a critical impact on model accuracy. We categorize the possible failures and propose hardware implementation that minimizes catastrophic failures. Such hardware optimization bounds the possible logical value of the failed cells and gives us opportunities to compensate the loss of accuracy via off-line training. By introducing the random weight defects during the training, we show that the model becomes more resilient on the device initialization failures, therefore, less prone to degrade the inference performance due to the failed devices. Our study shed light on the hardware and software co-optimization procedure to cope with potentially catastrophic failures in the cross-bar array. 
\end{abstract}



%
% Keywords. The author(s) should pick words that accurately describe the work being
% presented. Separate the keywords with commas.

%
% This command processes the author and affiliation and title information and builds
% the first part of the formatted document.
\maketitle
\thispagestyle{empty}

\section{Introduction} \label{sec:intro}

Recent progress in algorithm and computing hardware has made it possible to train the neural network in large scale and demonstrated that neuromorphic computing is a robust and efficient way to solve various problems including the pattern recognition \cite{he2016deep}, speech recognition \cite{zhang2016towards}, and optimization \cite{villarrubia2018artificial}. The central step of training is to modify the mapping rule from one neuron layer to the other adjacent layer minimizing the assumed cost function. Such mapping is often expressed as weight matrices and optimizing the weights requires intensive matrix operations setting a bottleneck in the training. 
While transistor scaling and the subsequent performance gain has resolved such bottleneck last few decades, the pace of the scaling has significantly slowed due to the growing cost and diminishing returns. The technological and economical challenges drive the computing more toward to the specialized computing architectures rather than the general purpose processors \cite{thompson2018decline}.   

Recently, cross-bar array hardware has been investigated as an emerging architecture. The architecture exploits the analog memory elements for multi-state weight representations and performs in-memory multiply-accumulate (MAC) operations \cite{gokmen2016acceleration, 2019Haensch}. The architecture is optimized to perform the MAC operations in parallel and such parallelism shows significant advantages over a conventional hardware both in speed and power consumption \cite{gokmen2016acceleration, chi2016prime,shafiee2016isaac, song2017pipelayer}. Moreover, the feasibility of the architecture has been demonstrated using non-volatile memory such as a phase change memory (PCM) \cite{ambrogio2018equivalent}, a resistive random access memory (ReRAM) \cite{bocquet2018inmemory,zhou2018anewhardware}, and charge trapping memory \cite{Merrikh2018,Guo2017} element.
 
As cross-bar array hardware has a fundamentally different physical realization from the conventional architecture, the optimal ways of implementing the cross-bar array in the architecture level has been investigated for the fully connected \cite{gokmen2016acceleration}, convolutional \cite{chi2016prime, shafiee2016isaac, song2017pipelayer, gokmen2018trainingConv}, and recurrent \cite{gokmen2018trainingLSTM, long2018reram} neural networks. The analog nature of the memory element, however, lacks the explicit quantization of the states and errors occurred in the analog hardware domain may be accumulative. Thus evaluating and understanding the impact of the non-idealities in the analog domain and the analog/digital interface is a key element to enable the cross-bar array technology. The device-to-device, cycle-to-cycle update variations \cite{gokmen2016acceleration} as well as the resistance drift \cite{liu2017analyzing} has been considered for the individual analog elements. The variations in the peripheral circuitry have been investigated including the error occurring at the analog/digital interface \cite{gokmen2016acceleration,gokmen2018trainingConv,gokmen2018trainingLSTM} and the sense amplify circuitry \cite{sun2018xnor}. In addition, general strategies to address the non-idealities in analog domain has been discussed \cite{Jain2019NeuralNA}.

Following this trajectory, we have investigated the possible device failure scenario and its impact on the model accuracy. The impact of the failed cell has been examined in the fully-connected neural network for the PCM based architecture \cite{romero2019training}. However, the lack of prior study on the ReRAM hardware motivates us to study the device failure scenario which occurs during the initialization process of the ReRAM based cross-bar array. After the fabrication of the device, the forming process is a necessary step to generate the filamentary conductive path for the ReRAM. During the forming process, the memory cell may stuck on the high-resistance or low-resistance state and potentially play as a critical source of error. Unlike the conventional memory, however, the analog memory elements are hard-wired and it is not straightforward to re-route the failed cells. Instead, one may incorporate a whole redundant arrays or columns to address the failed cells \cite{xia2017stuck}. To provide a different perspective, we attempt to address the failed cells in cross-bar array via hardware and software co-optimization without utilizing hardware redundancies. We first evaluate the potential impact of such failed cells on the architecture and propose the optimized hardware. We discuss on the inference model accuracy and propose an off-line training strategy that further compensates the accuracy lost due to the failed cells. 

\section{Modeling of initialization failure in ReRAM cross-bar array} \label{sec:model}
\begin{figure}[t!] 
  \centering
   \includegraphics[width=0.5\textwidth]{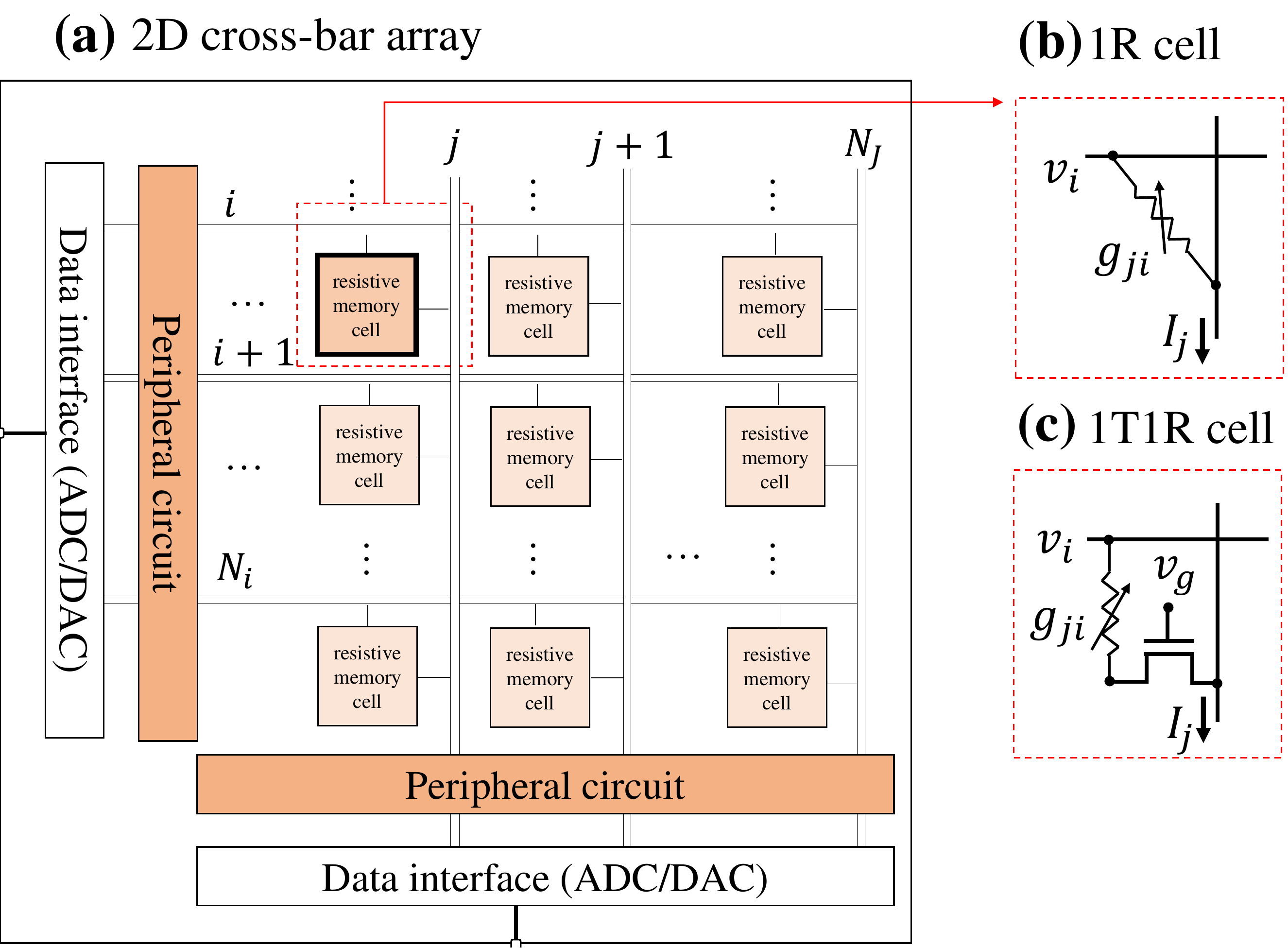}
  \caption{\textbf{The schematics of cross-bar array system.} (a) A cross-bar array is comprised of a resistive memory cell, peripheral circuits, and data interface. (b) The resistive memory cell stores weight values in a form of a resistance, which may consist of one resistor (1R). (c) The resistive memory cell may comprise one resistor and one transistor (1T1R) where the transistor plays a role to limit the current.}\label{fig:RPU}
\end{figure}

Using cross-bar array, we may perform multiplication and summation in parallel and improve calculation efficiency by a following mechanism. Figure~\ref{fig:RPU}(a) shows the individual resistive memory unit cell connected to a word-line (horizontal line) and a bit-line (vertical line). Assuming that the voltage across the word-line is $v_i$, the current read from the bit-line is $I_{ji}=g_{ji} v_i$, where $g_{ji}$ is the conductance of the resistive element. By applying a certain voltage to the available word-line at a given period of time $t_i\in[0,t_{max}]$ and integrate the output current, each bit-line reads the total accumulative charges as
\begin{equation} \label{eq:zj}
z_j=\sum_i \int_t dt I_{ji}=\sum_i w_{ji}\times a_i,
\end{equation}
where $a_i=v_it_{max}$ and $g_{ji}=w_{ji}$. Here, the conductance of the resistive element $g_{ji}$ represents the weight value $w_{ji}$ and the applied bias $v_i$ at the given word-line maps to the activation of the previous layer or an input value. As a result, Eq.~(\ref{eq:zj}) effectively represents the matrix multiplication of the hidden layer. Once the matrix multiplication is done by the cross-bar array, the results are processed in peripheral circuits and converted to the digital data at the data interface in Fig.~\ref{fig:RPU}(a) \cite{gokmen2016acceleration}. Then, other operations such as the batch normalization or activation functions are executed in the digital circuits before the next cross-bar array consumes the output data. For this reason, the weight elements utilized for the matrix multiplication are represented by the analog memory elements, whereas remainder weights including batch normalization parameters are assumed to be handled in the digital circuits in floating point precision.
%============FIGURE=============================
%==========FIGURE=================================

Among the various resistive memory candidates, we mainly focus on the resistive switching memory or resistive random access memory (ReRAM) whose resistance states are determined by a filamentary conductive path of a dielectric material. The filament is formed by applying large bias across the dielectric which induces a soft-breakdown of the material and this process is called \emph{forming}. For example, the breakdown process in HfO$_x$ occurs by the oxygen vacancy movement which creates a metallic conductive path \cite{kim2004engineering, pan2014recent}. Once such conductive path is \emph{formed}, the reverse polarity bias (or RESET bias) may be applied to induce the recombination of the oxygen and the oxygen vacancy. Once such recombination disconnects the filamentary conductive path, the device is in a high resistance state (HRS). Likewise, the same polarity with the forming bias (or SET bias) may be applied to re-connect the conductive filament to set the device in a low resistance state (LRS). Figure~\ref{fig:Grange} shows the ideal ReRAM cell which may allow us to explore intermediate states between LRS and HRS by applying SET and RESET bias. 
In this study, we assume that each ReRAM cell is written by write and verify method. This means that we assume the writing process involves in multiple checking and re-writing steps until we reach the desired state for each cell. Although such method becomes a serious bottle-neck of the algorithm for training purpose, it is a valid approach for inference purpose as we need to update the weight matrices once when we copy the trained model to the ReRAM cells. Using this approach, ReRAM cell have been demonstrated to express 7 bits (128 states) with $0.39\%$ fluctuations compared with the spacings between states \cite{li2018analogue}. Of course, there are other source of non-idealities such as noise and an intrinsic stochasticity in ReRAM. More comprehensive studies have been done, i.e. \cite{gokmen2016acceleration}, and have shown that the cross-bar array architecture is robust in certain degree to such non-idealities. Therefore, we assume that the ReRAM device that is not failed are ideal and the states are written exactly as it supposed to be up to the 4 bits of quantization levels.  

However, the device may fail during the forming process if (i) the ReRAM cell is not formed and remained open for a given forming bias. The forming voltage is a function of dielectric thickness and area \cite{chen2013area}. A ReRAM cell that has process variations on these parameters may result in the forming voltage higher than the maximum voltage that is supplied by the peripheral circuit, and may not be formed. In addition, (ii) the filament may be \emph{overformed} and cannot be disconnected by the RESET bias. To form the conduction path with a desirable filamentary thickness, it is important to control the current supplied during the forming process. The failure of the current control often results in overforming the device. A fast rate of filament formation (less than 1ns \cite{bersuker2019metal}) make it challenging to control the supplied current simply by removing the applied voltage timely. Therefore, a typical approach to limit the supply current is to utilize a transistor. However, the existence of the parasitic capacitance results in an overshoot of the current over the compliance current even in the presence of the transistor \cite{kinoshita2008reduction}. Such overshooting in current partially ascribe to form the filament thicker than the desirable value and may produce filament that is overformed and cannot be reset. Above mentioned failures result in the cell states stuck in (i) HRS or (ii) LRS. 

According to Eq.~(\ref{eq:zj}), an extreme resistance value results in an abnormally large weight value and may cause a non-trivial impact on the model accuracy. 
%Considering that the typical weight values of the neural network tend to converge into $0$, extreme values from the failed cell may have a non-trivial impact on the model accuracy. 
In conventional random access memory architecture, such failed cells may be re-routed to the working redundant cells. However, the cells in cross-bar array are hard-wired and the failed cells cannot be re-routed. The goal of this study is to evaluate an impact of the two major failure mechanism on the inference model, and provide methods to minimize the loss in the model accuracy as well as an insight on the acceptable forming yield.

\subsection{Individual resistive memory forming failure analysis} \label{sec:cellAnalysis}
%============FIGURE=============================
 \begin{figure}[t!] 
  \centering
   \includegraphics[width=0.5\textwidth]{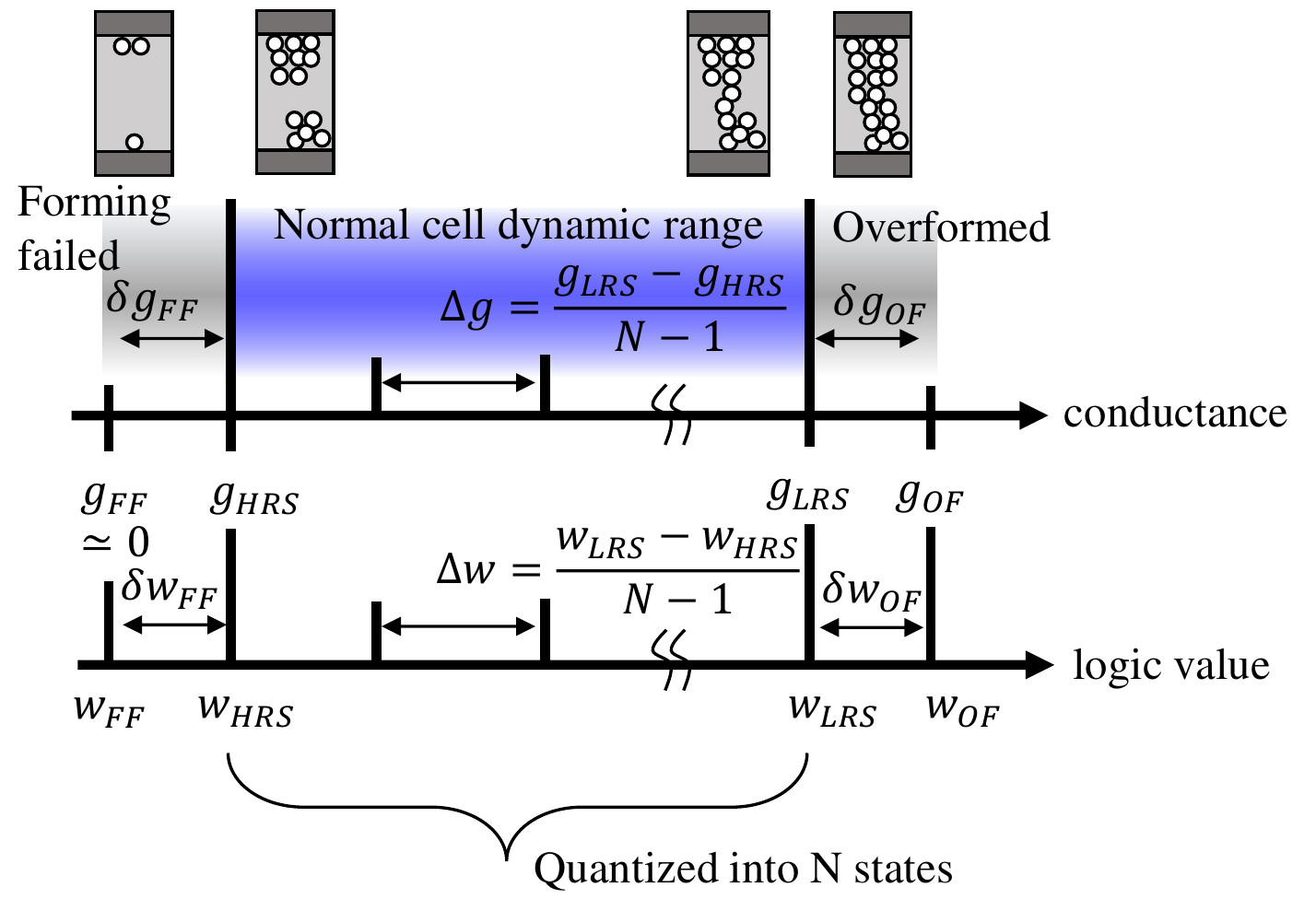}
  \caption{\textbf{The schematics of the conductance range of ReRAM.} The top schematics show the filament states and their corresponding conductance range. The normal cell dynamic range is defined by the conductance of high resistance state ($g_{HRS}$) and the low resistance state ($g_{LRS}$). The intermediate states are quantized into $N$ states and represents the relevant weight values ranging from $w_{HRS}$ to $w_{LRS}$. The conductance of the forming failed cell is close to $0$ and the resultant deviation from the $g_{HRS}$ is defined as $\delta g_{FF}$. Such deviation results in an additional error in weight value from $w_{HRS}$, which is defined as $\delta w_{FF}$. The overformed device tends to have lower resistance than the typical LRS. The resultant deviation from the desired LRS conductance is defined as $\delta g_{OF}$, and the corresponding deviation in the weight element is defined as $\delta w_{OF}$.}\label{fig:Grange}
\end{figure}   
%==========FIGURE=================================
Figure~\ref{fig:Grange} depicts the conductance range of the forming failed cells during the forming process. The top and bottom electrodes are electrically disconnected by the insulating dielectric for the forming failed (FF) devices. As a result, the conductance of the FF device typically shows few orders of magnitude lower than the conductance of the formed device. Therefore, we may assume the conductance of the FF devices as $g_{FF}\simeq0$. This assumption is consistent with the other works such as a dead device modeled as $g=0$ in PCM array \cite{romero2019training}, or a stuck-at-1 fail cell modeled as a minimum conductance (or $g\simeq0$) of the system in ReRAM array \cite{xia2017stuck}.
In contrast, the resistance of the overformed (OF) devices can be as low as few hundreds of $\Omega$ to few $k\Omega$ whereas the desirable ReRAM operation dynamic range is from few hundreds $k\Omega$ to few $M\Omega$ \cite{gokmen2016acceleration}. Assuming the dynamic range of the working device is $g\in[g_{HRS}, g_{LRS}]$, the corresponding logical value is mapped to $w\in[w_{HRS},w_{LRS}]$. However, if the conductance of OF or FF devices deviates significantly from a typical range of $g$, such scenario may lead to a serious failure of the cross-bar array architecture. For example, if the resistance states of the OF device stuck at $2~k\Omega$ whereas the LRS of the working cell is $100~k\Omega$, a single OF device flows current $50$ times larger than the expected LRS devices for a given input bias. In this case, the current level may be above the maximum acceptable range of the current integrator at the peripheral circuits and the whole bit-line signals may be overwhelmed. To prevent such scenario, we propose to use 1 transistor + 1 resistor (1T1R) cell structure shown in Fig.~\ref{fig:RPU}(c) rather than 1 resistor + 1 selector or 1 resistor (1R) structure described in Fig.~\ref{fig:RPU}(b). 1T1R structure limits the maximum current by adjusting the gate bias, $v_g$. Assuming that the transistor is operating in the linear regime, the drain current is determined as $I_{ds}=(W/L)\mu C_{ox}(v_{g}-v_{th})v_{ds}$, where $W$ and $L$ is the width and length of the gate, respectively, $\mu$ is the carrier mobility, $C_{ox}$ is the gate oxide capacitance, $v_{th}$ is the threshold voltage, and $v_{ds}$ is the bias across the source and drain. Therefore, we may set the effective conductance of the transistor $g_{tr}=(W/L)\mu C_{ox}(v_{g}-v_{th})$ close to the $g_{LRS}$ by adjusting the gate bias. In this case, the unit cell conductance range becomes
\begin{equation} \label{eq:glist}
\begin{split}
g_{LRS}^{cell}=&\frac{g_{LRS}\cdot g_{tr}}{g_{LRS}+g_{tr}}, \\
g_{HRS}^{cell}=&\frac{g_{HRS}\cdot g_{tr}}{g_{HRS}+g_{tr}}\simeq g_{HRS}, \\
g_{OF}^{cell}=&\frac{g_{OF}\cdot g_{tr}}{g_{OF}+g_{tr}}\simeq g_{tr}, \\
g_{FF}^{cell}=&\frac{g_{FF}\cdot g_{tr}}{g_{FF}+g_{tr}}\simeq g_{FF}, \\
\end{split}
\end{equation} 
where we assume $g_{FF}\ll g_{HRS}\ll g_{tr}\ll g_{OF}$. With the conductance value defined in Eq.~(\ref{eq:glist}), we may map the conductance to the logical value, or weight value. As it is described in Eq.~(\ref{eq:zj}), 
\begin{equation}
\begin{split}
g_{LRS}^{cell}\rightarrow& w_{LRS}, \\
g_{HRS}^{cell}\rightarrow& w_{HRS}, \\
\end{split}
\end{equation}
where $(w_{HRS},w_{LRS})\subset\{(0,2),(0,-1),(0,1)\}$ depending on the cross-bar array architecture. For the inference model, the weight value is often quantized into $N$ states without any significant accuracy loss of the model. Therefore, we may define the spacing of the conductance and the corresponding spacing in the logical value as 
\begin{equation} \label{eq:spacing}
\Delta g=\frac{g_{LRS}^{cell}-g_{HRS}^{cell}}{N-1}\rightarrow\Delta w=\frac{w_{LRS}-w_{HRS}}{N-1}.
\end{equation}
Note that the OF and FF devices are deviated from the expected maximum or minimum logical values, respectively. Figure~\ref{fig:Grange} describes such deviation as $\delta g_{FF}=g_{HRS}^{cell}-g_{FF}^{cell}$ for the FF device and $\delta g_{OF}=g_{OF}^{cell}-g_{LRS}^{cell}$ for the OF device. Their relative significance may be quantified by comparing with the logical value spacing defined in Eq.~(\ref{eq:spacing}):
\begin{equation} \label{eq:dwDw_min}
\begin{split}
\frac{\delta g_{FF}}{\Delta g}=\frac{\delta w_{FF}}{\Delta w}
=&(N-1)\frac{g_{HRS}^{cell}-g_{FF}^{cell}}{g_{LRS}^{cell}-g_{HRS}^{cell}} \\
\simeq&(N-1)\frac{1}{\bar{g}-1}, \\
\end{split}
\end{equation}
\begin{equation} \label{eq:dwDw_max}
\begin{split}
\frac{\delta g_{OF}}{\Delta g}=\frac{\delta w_{OF}}{\Delta w}
=&(N-1)\frac{g_{OF}^{cell}-g_{LRS}^{cell}}{g_{LRS}^{cell}-g_{HRS}^{cell}} \\
=&(N-1)\frac{g_{OF}^{cell}/g_{LRS}^{cell}-1}{\bar{g}-1} \\
\simeq&(N-1)\frac{g_{tr}}{g_{LRS}(\bar{g}-1)},
\end{split}
\end{equation}
where $\bar{g}=g_{LRS}^{cell}/g_{HRS}^{cell}$ is the min/max conductance ratio between LRS and HRS, and the last equalities in Eqs.~(\ref{eq:dwDw_min}-\ref{eq:dwDw_max}) are from Eq.~(\ref{eq:glist}). The logical value deviation of both OF and FF devices is minimized when the min/max ratio ($\bar{g}$) is maximized or the cell has less quantized states ($N$). In addition, setting $g_{tr}$ closer to the $g_{LRS}$ is beneficial in further reducing logical value deviation of the OF devices. However, this will reduce the overall dynamic range of the resistive unit cell due to the diminishing cell LRS conductance, $g_{LRS}^{cell}$, and eventually reduces $\bar{g}$. Prior to examine such trade-off, we need to further specify $N$. The quantized states $N$ is the number of states per individual cell, and its value may differ depending on the choice of the specific cross-bar array architecture. 

\subsection{Forming failure scenario analysis on resistive memory cell}
%============FIGURE=============================
 \begin{figure}[t!] 
  \centering
   \includegraphics[width=0.5\textwidth]{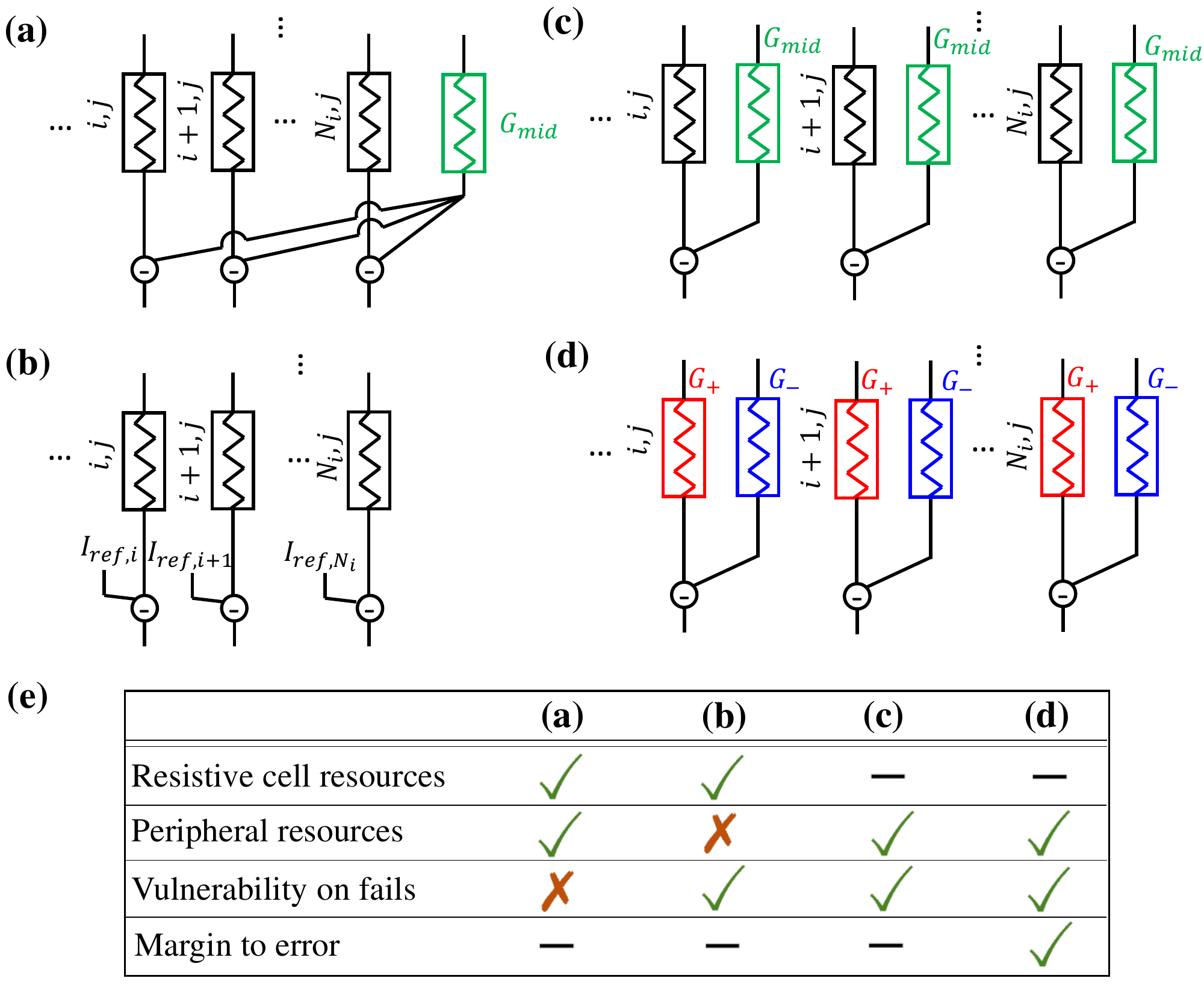}
  \caption{\textbf{Cross-bar array cell structure.} (a) Individual ReRAM cell represents $w_{ji}\in[0,2]$ and shares the reference column of ReRAM cells whose conductance is set to be $g_{mid}$, or $w_{mid}=1$. By subtracting the current from the reference column, each ReRAM cell represents full range of weight values, or $w_{ji}-w_{mid}\in[-1,1]$. (b) The reference current, $I_{ref}$, can be supplied from the external circuit. (c) Individual ReRAM cell is paired with the reference cell. (d) A pair of ReRAM cell represents plus ($w^+$) and minus ($w^-$) weight values. This choice of differential reading allows to reduce the number of quantized states per cell, as each cell needs to store half of the weight range. (e) Summary on the pros and cons of the cell structures. \xmark$\;$indicates an unacceptable disadvantages over other candidates, $-$ indicates no difference with others or an acceptable level of disadvantage. \checkmark$\;$indicates that the candidate is superior or equivalent to other candidates.}\label{fig:CellStructure}
\end{figure}   
%==========FIGURE=================================
Figure~\ref{fig:CellStructure} shows four possible resistive memory cell structures which are designed to express $w_{ji}\in[-1,1]$ logical values. The most efficient implementation in hardware resource perspective may be shown in Fig.~\ref{fig:CellStructure}(a). In this structure, each individual cell represents $(w_{min},w_{max})=(0,2)$ and the cells in the last column serve as reference devices by fixing its conductance value to $g_{mid}=(g_{max}+g_{min})/2$, or $w_{mid}=1$. By subtracting the individual cells with the reference column, the architecture represents plus and minus values of the weights, or $w_{ji}-w_{mid}\in[-1,1]$. However, the impact of the OF or FF device exerts on the whole row if a forming process failure occurs to the reference device. To avoid such scenario, we may simply subtract current at the end of the column-wise current integration as described in Fig.~\ref{fig:CellStructure}(b) by introducing $I_{ref,i}=\sum_i v_i\cdot g_{mid}$. While this resolves the problem, we now need to calculate $I_{ref,i}$ using extra-peripheral circuits. Instead, Fig.~\ref{fig:CellStructure}(c) shows that we may double the hardware resources by introducing an additional resistive memory to form a resistive memory unit cell. Here, we use one cell as a weight storage and the other cell as a reference device. Note that each weight storage cell represents $(w_{min},w_{max})=(0,2)$, which is quantized into $N=2^{n_{bit}}$ levels, where $n_{bit}$ is the target model quantization bit. An alternative scheme in Fig.~\ref{fig:CellStructure}(d) utilizes one cell for a plus logical value, or $(w_{min}^+,w_{max}^+)=(0,1)$, and the other cell for a minus logical value, or $(w_{min}^-,w_{max}^-)=(-1,0)$. As the $n_{bit}$ quantization is performed over $(-1,1)$, each cell now stores $N=2^{n_{bit}-1}$ quantized levels as each of the cell covers half of the weight range. According to the Eqs.~(\ref{eq:dwDw_min}-\ref{eq:dwDw_max}), the 2T2R structure (two 1T1R) in Fig.~\ref{fig:CellStructure}(d) is advantageous over other candidates as it reduces the number of quantized states from $N=2^{n_{bit}}$ to $N=2^{n_{bit}-1}$ which effectively reduces the logical value deviation of OF and FF devices. The above mentioned arguments are summarized in Fig.~\ref{fig:CellStructure}(e).

%============FIGURE=============================
 \begin{figure}[t!] 
  \centering
   \includegraphics[width=0.5\textwidth]{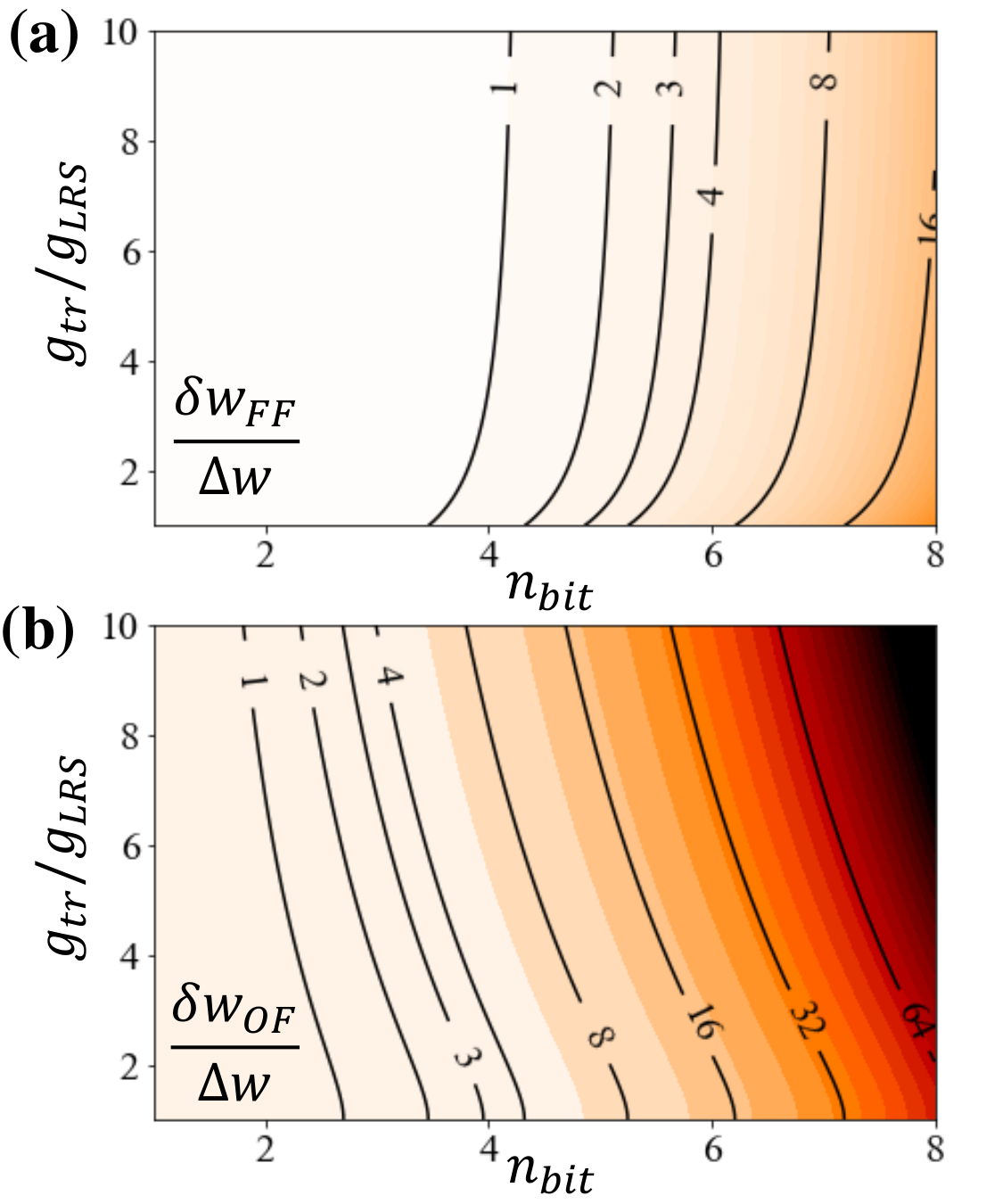}
  \caption{\textbf{A logical value deviation of FF and OF devices.} (a) A logical value deviation of FF device ($\delta w_{FF}$) as a function of a quantization bit ($n_{bit}$) and conductance of the transistor ($g_{tr}$). $\delta w_{FF}$ is calculated from Eq.~(\ref{eq:dwDw_min}). The contour line shows $\delta w_{FF}/\Delta w=1,\;2,\;3,\;4,\;8$ and $16$. (b) A logical value deviation of OF device ($\delta w_{OF}$). $\delta w_{OF}$ is calculated from Eq.~(\ref{eq:dwDw_max}). The contour line shows $\delta w_{OF}/\Delta w=1,\;2,\;3,\;4,\;8,\;16,\;32$ and $64$. We set the min/max ratio of the conductance $g_{LRS}/g_{HRS}=10$ for (a) and (b).}\label{fig:wFFwOF}
\end{figure}   
%==========FIGURE=================================
Adopting the 2T2R resistive memory cell described in Fig.~\ref{fig:CellStructure}(d), we examine how undesirable conductance deviation in Fig.~\ref{fig:Grange} is affected by the physical parameter variations and quantify its significance in terms of the weight quantization step. Specifically, we vary the number of quantized conductance ($n_{bit}$) and the effective gate conductance ($g_{tr}$). Figure~\ref{fig:wFFwOF}(a) and (b) shows the resultant logical weight deviation $\delta w_{FF}$ and $\delta w_{OF}$, respectively, with their magnitude normalized by the quantized step ($\Delta w$).  
We choose to vary $n_{bit}$ and $g_{tr}$ for the following reasons: (i) Eqs.~(\ref{eq:dwDw_min}-\ref{eq:dwDw_max}) shows that these variables determine $\delta w_{FF}$ and $\delta w_{OF}$ and (ii) they are tunable parameters without any significant modification in the hardware. If we lower the quantization bit, the quantization step $\Delta w$ is now represented by the larger conductance window. This wider window allows the cell to tolerate a given amount of the conductance deviation. As a result, smaller $n_{bit}$ always improves the logical error of the resistive memory cell. Figure~\ref{fig:wFFwOF}(a) shows that the logical error of FF device approaches to one quantization step when $n_{bit}\sim4$. For a given $n_{bit}$, larger $g_{tr}$ helps to reduce $\delta w_{FF}$. This is due to the fact that larger transistor conductance allows a larger dynamic range of the resistive cell, thus, the relative significance of the conductance deviation is reduced. In contrast, Fig.~\ref{fig:wFFwOF}(b) shows that smaller $g_{tr}$ helps to reduce $\delta w_{OF}$. In this case, having $g_{tr}$ closer to $g_{LRS}$ sets a better conductance lower-bound for OF devices. The overall trade-off shows that optimizing the gate bias to adjust smaller $g_{tr}$ is desirable in reducing the logical weight value deviations rather than fully turn-on the transistor during the MAC operation. This is because $\delta w_{OF}$ is more sensitive to the parameters than $\delta w_{FF}$. For example, $\delta w_{OF}$ is as large as three quantization steps at $(n_{bit},g_{tr}/g_{LRS})=(4,1)$, whereas $\delta w_{FF}$ is equivalent or smaller than one quantization step in wider range of parameters. 
Note that we fix the min/max ratio of the resistive cell as $g_{LRS}/g_{HRS}=10$ in Fig.~\ref{fig:wFFwOF}. The logical errors are further minimized for higher min/max ratio, although improving min/max ratio requires longer term effort as it requires an improvement in materials and device structures. %Figs.~\ref{fig:wFFwOF}(a-b) shows that the logical error induced by OF devices is more severe than the error induced by FF devices.  

\begin{table}[t!]
\caption{Summary of the possible forming failure types and the resultant logical error and its occurrence probability. Working cells are indicated as -.} % title of Table
\centering % used for centering table
\begin{tabular}{c c p{1.5cm} p{1.5cm} p{1.5cm} c} % centered columns (4 columns)
\hline\hline %inserts double horizontal lines
$w^+$ & $w^-$ & possible value range & best possible value & logical error & prob. \\ [0.5ex] % inserts table
%heading
\hline % inserts single horizontal line
FF & -  & $[-1,0]$ & $0$ & $-\delta w_{FF}$ & $p_{FF}(1-p)$ \\ % inserting body of the table
FF & FF  & $0$ & $0$ & $\sim 0$ & $p_{FF}p_{FF}$ \\
FF & OF & $-1$ & $-1$ & $-\delta w_{OF}$ & $p_{FF}p_{OF}$ \\ 
OF & - & $[0,+1]$ & $0$ & $+\delta w_{OF}$ & $p_{OF}(1-p)$ \\
OF & FF  & $+1$ & $+1$ & $+\delta w_{OF}$ & $p_{OF}p_{FF}$ \\
OF & OF & $0$ & $0$ & $\sim0$ & $p_{OF}p_{OF}$ \\ [1ex] % [1ex] adds vertical space
- & - & $[-1,+1]$ &  & 0 & $(1-p)(1-p)$ \\
- & FF  & $[0,+1]$ & $0$ & $+\delta w_{FF}$ & $(1-p)p_{FF}$ \\
- & OF & $[-1,0]$ & $0$ & $-\delta w_{OF}$ & $(1-p)p_{OF}$ \\ [1ex] % [1ex] adds vertical space
\hline %inserts single line
\end{tabular}
\label{tb:summary} % is used to refer this table in the text
\end{table}

Even with a zero logical value deviation, the conductance state of the OF device is fixed near $g_{max}^{cell}$. Therefore, the corresponding logical value is stuck at $\pm1$. We define such failed cells as $\pm1$ defects in the weight matrices. In contrast, the conductance of the FF device is close to $g_{min}^{cell}$ and the resultant logical value is stuck at $0$. We define these failed cells as $0$ defects. 
With 2T2R structure in Fig.~\ref{fig:CellStructure}(d), we further minimize $\pm1$ defects by utilizing the following forming protocol. For example, in case the $w^+$ cell is overformed while its pair $w^-$ is properly working, we can avoid having unintentional $+1$ value by setting $w^-=-1$. In this case, the conductance of the OF device in $w^+$ is compensated by the conductance of $w^-$ and we have $0+\delta w_{OF}$ as a logical error instead of $+1+\delta w_{OF}$. In other words, by utilizing the resistive cell pairs, we effectively prune the mal-functioning weight elements instead of having $+1$ logical error. Table~\ref{tb:summary} summarizes all the possible scenario. In Table~\ref{tb:summary}, we define the probability for FF and OF devices as $p_{FF}$ and $p_{OF}$, respectively, and the total forming failure probability is defined as $p=p_{FF}+p_{OF}$. The straightforward initialization protocol (referred to as a strategy A) may faithfully follow the Table~\ref{tb:summary}. In this case, the resistive memory cells stuck at $\pm1$ value with a probability of $p_1=2p_{FF}p_{OF}$ while $(1-p)^2$ cells are properly working. The remaining portion of the cell is forced to be $0$ and the weights are effectively pruned to minimize the impact of the failed cells. The probability of having $0$ defect is $p_0=2(1-p)p+p_{OF}p_{OF}+p_{FF}p_{FF}\simeq 2p$ if $p\ll 1$. Among all possible scenarios for $0$ defects, $(w^+,w^-)=(FF,-),\;(-,OF)$ and $(w^+,w^-)=(-,FF),\;(OF,-)$ cases are programmable and may programmed correctly if the weight value is negative or positive, respectively. If we assume that the chance of having minus or plus weight value for an arbitrary cell is $50\%$, the resistive cell still records the correct value with the probability of $p_0\simeq2p/2\simeq p$. As a result, we end up having $0$ defects with a probability of $\sim p$. 

\begin{table}[t!]
\caption{Two forming strategies based on Table~\ref{tb:summary} and the resultant $0$ defect ($p_0$) and $\pm1$ defect ($p_1$) probabilities.} % title of Table
\centering % used for centering table
\begin{tabular}{c c c} % centered columns (4 columns)
\hline\hline %inserts double horizontal lines
 & strategy A& strategy B \\ [0.5ex] % inserts table
%heading
\hline % inserts single horizontal line
$p_0$ & $\sim p$ & $\sim2p$ \\ % inserting body of the table
$p_1$ & $2p_{OF}\cdot p_{FF}$ & $p_{OF}\cdot p_{FF}$ \\
\hline %inserts single line
\end{tabular}
\label{tb:summary2} % is used to refer this table in the text
\end{table}
Another possible strategy (referred to as strategy B) is to form one device first and choose not to form its pair if the device is FF device. This strategy will avoid the risk of having OF device for its pair during the forming process, and further reduce the probability of having $\pm1$ defect from $2p_{FF}p_{OF}$ to $p_{FF}p_{OF}$. Although such choice is effective in minimizing $\pm1$ defects, we no longer have $(w^+,w^-)=(FF,-)$ or $(-,FF)$ as we choose not to form the pair of FF devices. If we simply assume that we do not write the weight values to the failed cells, the probability of having $0$ defect is now $p_0\simeq 2p$. 

When the OF device occurrence is low, it is more beneficial to choose the strategy A due to the lower $p_0$ probability. However, the strategy B is a good option if OF device occurrence rate is substantial compared with the occurrence of the FF device. The discussed two forming strategy A and B and their corresponding probabilities of having $\pm1$ defect ($p_1$) and $0$ defect ($p_0$) are summarized in Table~\ref{tb:summary2}. The numerical analysis in the following section utilizes the strategy B, but the qualitative results are consistent for both strategy A and B as they differs only by the probability combinations of $0$ and $\pm1$ defects.

\section{Numerical experiments}

\subsection{ The impact of $\pm1$ and $0$ defects on the inference accuracy}\label{sec:defectAware}
We use CIFAR-10 dataset and test the impact of the forming failed cells on the image recognition model. The model has been trained in ResNet-20 \cite{he2016identity}. 50,000 images has been used for training and 10,000 images are utilized for the test. According to the analysis in Section~\ref{sec:cellAnalysis}, a deviation of the weight from the desired logical value is minimized for smaller quantized bit, or $n_{bit}$. We quantize the weight \cite{zhou2016dorefa} and activation \cite{choi2018pact} and find no significant degradation in the test accuracy for $n_{bit}\geq4$ both for weight and activation \cite{choi2018pact, distiller2018}. Below $n_{bit}=4$ requires special techniques to maintain the model accuracy \cite{rastegari2016xnor, courbariaux2016binarized} which is out of scope in this study, thus, we choose $n_{bit}=4$. We obtain the baseline with a test error of $7.99\%$ after 200 epochs of training with a learning rate scheduling of $0.1$, $0.01$, $0.005$ at $80$, $120$, $180$ epochs. 

%============FIGURE=============================
 \begin{figure*}
  \centering
   \includegraphics[width=1.0\textwidth]{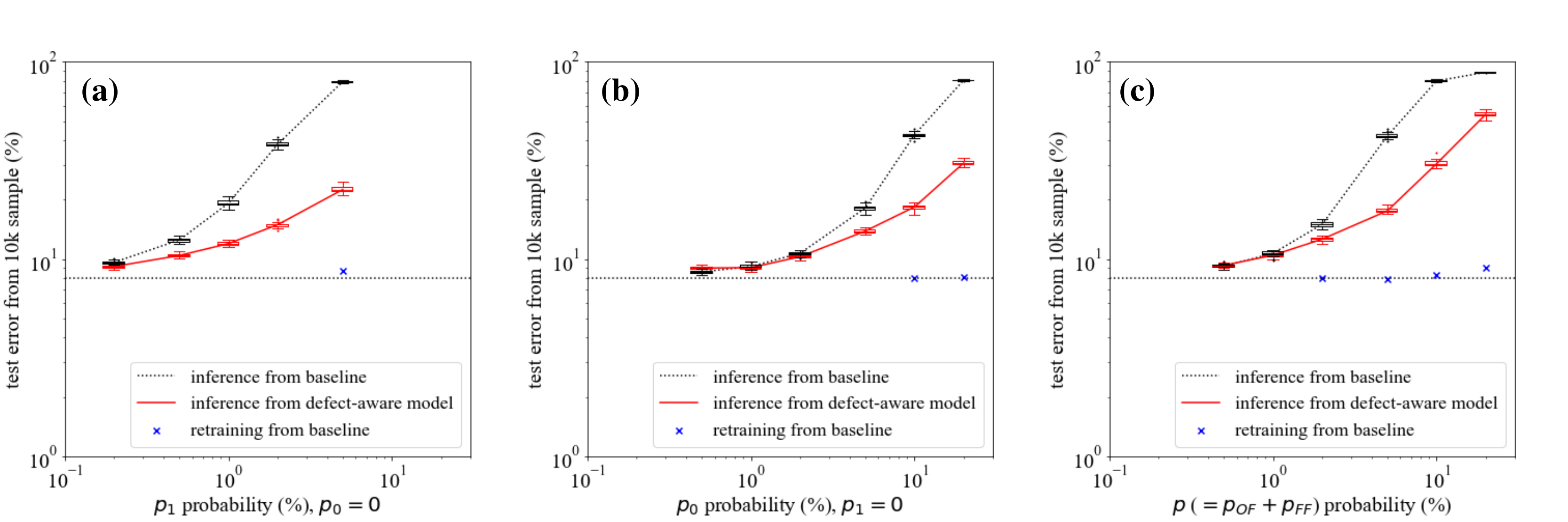}
  \caption{\textbf{Inference error in the presence of $\pm1$ and $0$ defects.} Black dotted line shows the inference error from the baseline, and red solid line shows the inference results from the defect-aware model. Each box plot represents the inference results from $50$ different defect configurations. The blue cross symbol shows the retraining results from one particular random defect configuration. (a) The impact of the $\pm1$ defects on the inference accuracy. The weight elements of the trained model are replaced by $\pm1$ value with a give probability $p_1$. (b) The impact of the $0$ defects on the inference accuracy. The weight elements of the trained model are replaced by $0$ value with a give probability $p_0$. (c) The impact of both $\pm1$ and $0$ defects on the inference accuracy. We vary the probability of forming failure $p=p_{OF}+p_{FF}$, and the corresponding $\pm1$ and $0$ defects are placed with the calculated probability following Table~\ref{tb:summary}. 
  }\label{fig:experiment}
\end{figure*}
%==========FIGURE=================================

We first assume an ideal defect by setting $\delta w_{OF,FF}\rightarrow 0$ and evaluate the impact of $0$ and $\pm1$ random defects on the trained model. Specifically, we randomly change the weight matrix elements of the trained model to $0$ or $\pm1$ with a probability of $p_0$ or $p_1$, respectively, and we set an equal probability for $+1$ and $-1$ defects for simplicity. As a first step, we isolate the impact of the $0$ defects from $\pm1$ defects by setting $p_0=0$. We then vary $p_1$ and obtain the inference results. The black dashed line in Fig.~\ref{fig:experiment}(a) shows the inference results for $p_1=0.2\%-5\%$. The test error rapidly increases and the model shows more than $20\%$ error rate for $p_1>1\%$. However, such degradation in the test accuracy may be recovered by retraining the network if we know the exact defect configuration. The blue x symbol in Fig.~\ref{fig:experiment}(a) shows the recovered accuracy through 200 epochs of re-training from the baseline model using a fixed defect configuration and an identical learning rate scheduling with the baseline. This approach is valid when we know the exact location of the existing defects. The defect configuration, however, likely appears in random and varies from chip to chip. Therefore, the strategy requires a prior knowledge of a specific configuration as well as computational resources to re-train for each individual chips. 

Instead, we may train the network with a prior knowledge of a type of defect and its likelihood. By introducing random defects for a given probability during the training, the model may find a local minimum which minimizes the loss function even in the presence of random defects. Through this process, model may become resilient to a particular set of defects without a knowledge of a specific defect configuration, thus a single trained model may be utilized for numerous defect configurations. We examine the hypothesis by introducing randomly generated $\pm1$ defects with a probability $p_1$. After the weight matrices at each layer are quantized, certain portion of the weight elements are replaced by $+1$ in $p_1/2$ and $-1$ in $p_1/2$ probability. The modified weight matrices are utilized for a given training image to perform the forward, and backward propagation. We then repeat the above procedure by generating a new set of random defects in the same probability for the next training image and the defect-aware model has been trained with 200 epochs. Using the resultant model, we generate $50$ different random defect configurations. The corresponding inference results are plotted as a red solid line in Fig.~\ref{fig:experiment}(a). When the defect probability is smaller than $0.2\%$, the inference results from baseline (black dotted line) and defect-aware model (red line) shows little difference. However, as we increase the defect probability, the defect-aware model produces much better inference results than the inference results from the baseline. 
%In fact, introducing $0$ defect is equivalent to randomly pruning neurons at a given layer. The accuracy of the randomly pruned model has known to be recovered by retraining the model \cite{mittal2018recovering}. The blue x symbol in Fig.~\ref{fig:experiment}(b) depicts that the model accuracy has been recovered close to the baseline after retraining.

We perform the similar analysis for the $0$ defects. Namely, we replace the quantized weight matrix elements to $0$ with a probability of $p_0$. Figure~\ref{fig:experiment}(b) shows a consistent trend with Fig.~\ref{fig:experiment}(a). One noticeable difference is the overall shift of the model error degradation. The model accuracy shows a significant degradation when $p_0>\sim1\%$ for $0$ defect whereas $p_1>\sim0.2\%$ for $\pm1$ defect. In other words, $\pm1$ defect exhibits more critical impact on the model accuracy. This may be understood by looking at the weight value distribution. After the 200 epoch training, most of the baseline weight values are centered at $0$ as the loss function tends to minimize the weight value. The test error is expected to increase upon an introduction of the defects as the difference between the desired weight value and the defect plays as a source of error. The model weight matrices have much less $\pm1$ values than $0$, therefore, an arbitrary change of the weight value to $\pm1$ may induce larger amount of error in higher chance. Through the defect-aware training, however, the weight value distribution is forced to shift toward $\pm1$ by intentionally populating extreme weight values in random fashion. Although the resultant model is deviated from the global minimum, the model becomes less prone to fail when similar type of defects are introduced.  

To keep the test error of CIFAR-10 dataset below $10\%$, for example, the baseline results shows that one needs to keep the $\pm1$ defect probability less than $0.1\%$, whereas the defect-aware model may tolerate up to $\sim0.5\%$. In contrast, the baseline model is less susceptible to $0$ defects. As the baseline model inherently includes certain amount of randomly distributed weight values close to $0$, both baseline and defect-aware model tolerates the defect probability up to $\sim 1\%$. The baseline and defect-aware model start to show a noticeable difference once we introduce more than $2\%$ of the $0$ defects. This summarizes that the defect-aware model approach is effective when the defect type tends to manifest itself as an uncommon logical value from typical weight value distributions. 
Furthermore, Fig.~\ref{fig:experiment}(a-b) shows that the baseline model is more sensitive to $\pm1$ defects. This result further justifies the 2T2R structure over 1T1R shown in Fig.~\ref{fig:CellStructure}(d) as it reduces the probability of having $\pm1$ from $p_{OF}$ to $p_{OF}p_{FF}$. 

With our understanding on the impact of $\pm1$ and $0$ defects on the inference model accuracy, we now utilize the Table~\ref{tb:summary} to evaluate the optimized hardware performance as a function of the probability of having OF and FF devices. When the total probability of the initialization failure scenario is $p=p_{OF}+p_{FF}$, we choose $p_{OF}=p_{FF}$ which is the worst case scenario of having maximum $\pm1$ defect. Figure~\ref{fig:experiment}(c) shows the inference results as a function of $p$. The 2T2R hardware minimizes the $\pm1$ defects and the inference results get close to the results of the $0$ defect in Fig.~\ref{fig:experiment}(b).

\subsection{The impact of $\delta w_{OF,FF}$ on the inference accuracy}
%============FIGURE=============================
 \begin{figure}[t!] 
  \centering
   \includegraphics[width=0.5\textwidth]{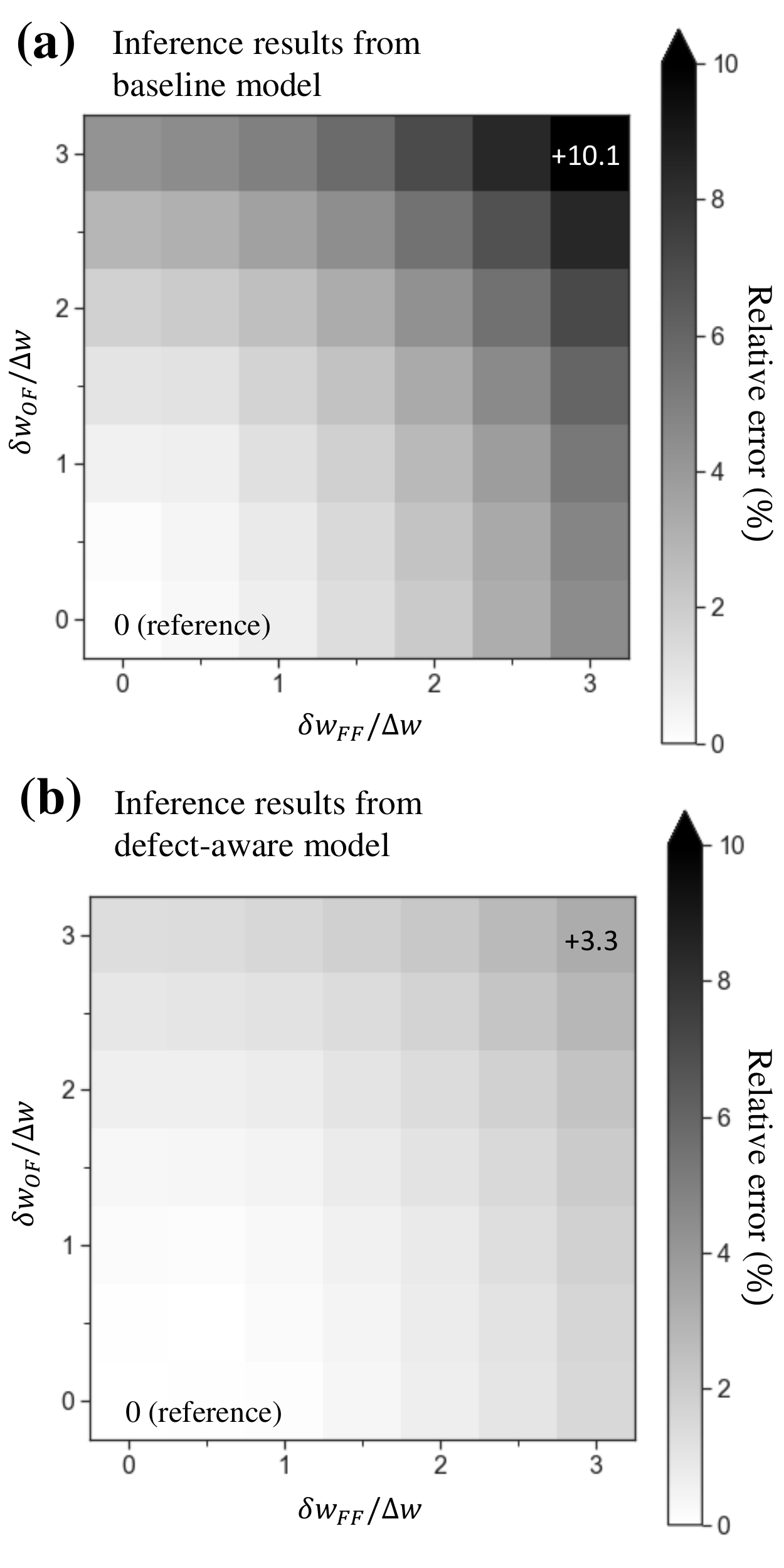}
  \caption{\textbf{Robustness of the defect-aware model to the non-ideality of the defects ($\delta w_{OF, FF}$) on the test error.} (a) The inference error from the baseline at $p=2\%$ is $\sim16.2\%$ with $\delta w_{OF, FF}=0$. The plot shows the relative error increase caused by the non-zero $\delta w_{OF, FF}$. (b) The inference error from the defect-aware model at $p=2\%$ is $\sim12.8\%$ with $\delta w_{OF, FF}=0$. The plot shows the relative error increase by $\delta w_{OF,FF}$ is smaller than the results from the baseline model. }\label{fig:nonideal}
\end{figure}   
%==========FIGURE=================================
In reality, the conductance value of the OF device may be higher than the normal cells and, for example, the logical value becomes $+1+\delta w_{OF}$ instead of $+1$. Similarly, the conductance of the FF device may exhibit lower conductance than the expected value which results in $0\pm\delta w_{FF}$. Such non-ideal deviation from the expected logical value has been described in Fig.~\ref{fig:Grange} and formulated in Eqs.~(\ref{eq:dwDw_min}-\ref{eq:dwDw_max}). In our cell design, the possibility of having $\pm1$ or $0$ defects and the corresponding non-ideal deviation has been summarized in Table~\ref{tb:summary}. Following the prescriptions in the table, we examine the impact of such non-ideal deviations of the logical value of $\pm1$ and $0$ defects for $\delta w_{FF}, \delta w_{OF}$. Figure~\ref{fig:wFFwOF} shows that $\delta w_{FF}/\Delta w\sim1-2$ and $\delta w_{FF}/\Delta w\sim3-4$ are reasonable range for $n_{bit}=4$ and $g_{LRS}/g_{HRS}=10$. Therefore, we choose a range of parameters $\delta w_{FF},\;\delta w_{OF}\in[0, 3\Delta w]$, where $\Delta w=(1-(-1))/(2^{n_{bit}}-1)$ is the quantization step of the weight elements. 

Figure~\ref{fig:nonideal}(a) shows the inference results from the baseline for $p=p_{OF}+p_{FF}=2\%$. The inference error is averaged over $15$ different defect configurations and the error is $16.2\%$ for $\delta w_{FF}=\delta w_{OF}=0$. As we increase the non-ideal deviation of FF and OF devices, the averaged inference error is computed over $15$ different defect configurations and a relative error with respect to the error at $\delta w_{FF}=\delta w_{OF}=0$ is plotted in Fig.~\ref{fig:nonideal}(a). The error monotonically increases as $\delta w_{FF}$ and $\delta w_{OF}$ increase and we lose additional $\sim10.1\%$ accuracy for $\delta w_{FF}=\delta w_{OF}=3\Delta w$.

We perform the similar analysis on the defect-aware model for $p=p_{OF}+p_{FF}=2\%$ and the inference error has been obtained by averaging over results from $15$ different defect configurations. We first obtain the inference error of $12.8\%$ for $\delta w_{FF}=\delta w_{OF}=0$, which shows a less error than the inference results from the baseline. As we increase $\delta w_{FF}$ and $\delta w_{OF}$, the monotonically increasing error shows a consistent trend with that of the baseline inference results. However, the relative error is $\sim3.3\%$ at $\delta w_{FF}=\delta w_{OF}=3\Delta w$, which is three times smaller than the relative error observed in the baseline results. Therefore, the defect-aware model shows more robust inference capability against non-ideal deviations of the weight elements. 

\subsection{The impact of the defect probability distribution on the inference accuracy}
%============FIGURE=============================
 \begin{figure}[t!] 
  \centering
   \includegraphics[width=0.5\textwidth]{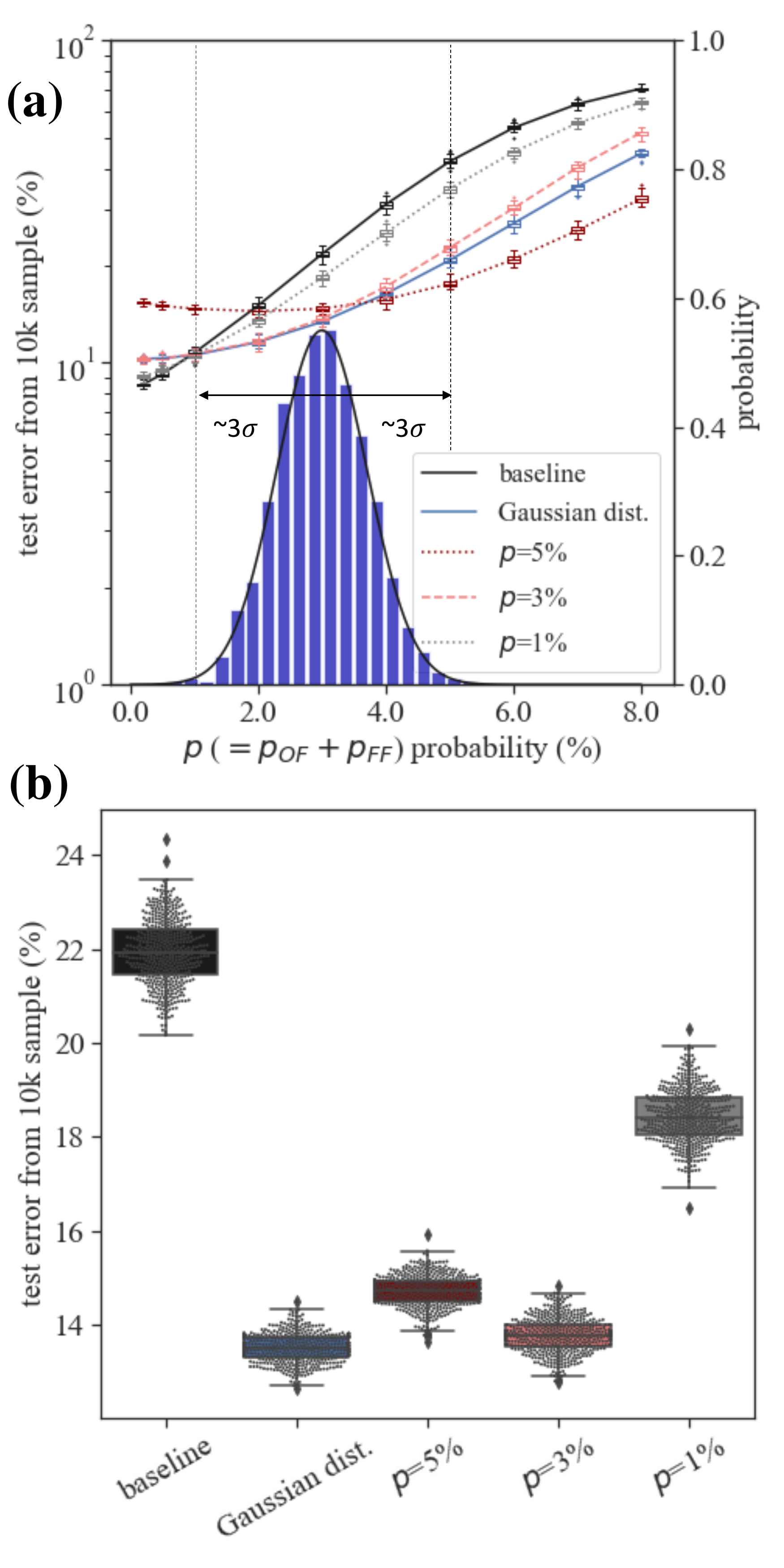}
  \caption{\textbf{Robustness of defect-aware model to the impact of the statistical variation of failure probability.} (a) Inference results from the baseline model (black solid line), defect-aware model (dashed line), and distribution-aware model (blue solid line). Each box plot represents the inference results from the $50$ different defect configuration. The left $y$-axis shows the assumed Gaussian distribution with a mean defect probability $\mu_p=3\%$ and a standard deviation $\sigma_p\simeq 0.7\%$. (b) Inference results from the same set of models in (a), but utilizes the assumed defect probability distribution shown in (a). The black-dot in box plot represents a possible inference test error for individual inference chip and their possible mean and standard deviation is presented in a box plot.}\label{fig:gaussian}
\end{figure}   
%==========FIGURE=================================
We may extract the initialization failure probability by collecting the statistics of the individual ReRAM data for a given die. However, such statistics may vary from die to die due to the process variation across the wafer, or from wafer to wafer due to the process drift induced by the equipment. As we have only discussed the defect-aware model at a fixed probability so far, it is worthwhile to investigate the strategy to address the statistical variations in the defect probability. For a given statistically meaningful interval of $p$, we pursue to obtain lower test error with minimal standard deviation. We proceed our discussion with an exemplary Gaussian distribution of the defect probability. 

The failure analysis on 4Mb ReRAM test chip shows $p_{OF}\sim1.75\%$ and $p_{FF}\sim9.04\%$ \cite{chen2014rram}.
After applying a dc bias to form 128Kb arrays, applying alternating set/reset pulses results in $p_{OF}\sim1.28\%$ and $p_{FF}\sim4.76\%$ \cite{shih2011training}. A series of optimized pulse inputs has been applied to 4Kb arrays and showed the improved forming yield, yet to have $p_{FF}\sim1\%$ \cite{grossi2016electrical}. 
Based on these results, we set a reasonable distribution of $p_{OF}$ and $p_{FF}$ to illustrate our approach. Specifically, we assume the mean values of $\mu_{p_{OF}}=\mu_{p_{FF}}=1.5\%$ with standard deviations of $\sigma_{p_{OF}}=\sigma_{p_{FF}}=0.5\%$. As a result, the total defect probability has a mean value of $\mu_p=\mu_{p_{OF}}+\mu_{p_{FF}}=3\%$ and $\sigma_{p}=\sqrt{\sigma_{p_{OF}}^2+\sigma_{p_{FF}}^2}\simeq 0.7\%$. The right-side $y$-axis of Fig.~\ref{fig:gaussian}(a) describes the statistics of the defect probability and a relevant probability interval from $p=1\%$ ($-3\sigma_p$) to $p=5\%$ ($+3\sigma_p$) has been indicated in dashed vertical lines.
 
The black line in Fig.~\ref{fig:gaussian}(a) and the left-side $y$-axis depict the inference results from the baseline model. A rapid increase of the test error has been observed from $\sim 10\%$ at $p=1\%$ to $\sim 40\%$ at $p=5\%$. The result shows that the baseline model is sensitive to the statistical variation of the defects due to its steep slope. The observed variation of the test error is unacceptably large and, therefore, a small deviation from the expected defect probability may result in a large deviation from an expected inference accuracy. The remaining plot shows the improved results from the defect-aware model. The gray dotted line in Fig.~\ref{fig:gaussian}(a) shows the inference error from the defect-aware model trained at $p=1\%$ ($-3\sigma_p$). A minor improvement over the baseline has been observed, but still suffers a rapid degradation of the accuracy for larger $p$. The dark red dotted line in Fig.~\ref{fig:gaussian}(a) exhibits the inference results of the defect-aware model trained at $p=5\%$ ($+3\sigma_p$). %Although the model shows an improved accuracy for $p=5\%$, the excessive amount of defects results in less accurate prediction for smaller defect probabilities. 
As its test error saturates for $p<3.0\%$, the variation of the inference error becomes less sensitive to the statistical variation. However, the overall test error is even higher at $p<2.0\%$ than the baseline results, which results in an overall degradation of the averaged test accuracy. The pink dashed line in Fig.~\ref{fig:gaussian}(a) shows the inference error from the defect-aware model at $p=3\%$. The model shows a better accuracy than the results from the baseline or the defect-aware model trained at $p=1\%$ within the relevant range of $p$. The inference results at $p>4\%$ shows an inferior performance than the model with $p=5\%$, but the statistical performance is expected to be better as the test error is comparable or smaller within the most significant interval of $\mu_p-\sigma_p\leq p \leq\mu_p+\sigma_p$. 

Alternatively, we may utilize the known distribution of the defect probability while the training. We randomly select $p_{OF}$ and $p_{FF}$ based on the known distribution and generate a defect configuration. The generated defect configuration is utilized for forward and backward propagation for a given image during the training procedure. The same procedure is repeated to generate a new configuration with a different defect probability for the next image training cycle. The inference results from the distribution-aware model is indicated as a blue solid line in Fig.~\ref{fig:gaussian}(a). The test error of the distribution-aware model shows comparable or less test error compared with the defect-aware model trained at the mean value of $p=3\%$ and, therefore, is expected to perform better in terms of mean and standard deviation of the inference test error.

Figure~\ref{fig:gaussian}(b) shows the inference results distribution from various models discussed in Fig.~\ref{fig:gaussian}(a). The defect probability $p_{OF}$ and $p_{FF}$ are randomly selected from the assumed distribution and we generate defect configurations with the selected probability. The same procedure is repeated for $500$ different trials. Each point represents one inference chip whose defect probability follows a given Gaussian distribution and the resultant statistical inference performances are presented. Figure~\ref{fig:gaussian}(b) clearly shows that both mean and standard deviation is the lowest for the distribution-aware model whose test error is $\mu\pm\sigma=13.5\pm 0.3\%$. 

\section{Summary and conclusion}
We have investigated the failure mechanism of the forming process in the resistive memory based cross-bar array. As the forming process is an essential initialization step for ReRAM devices, it is pivotal to understand the impact of the forming failure on the cross-bar array performance. Specifically, we have focused on the two failure scenario: (i) forming failure (FF) devices and (ii) overformed (OF) devices. In the cross-bar array architecture, the FF device flows minimal current and may act as $0$ defects whereas OF device allows an excessive amount of current which may cause a catastrophic error. We have discussed 1T1R structure to regulate the excessive current flow and set a lower bound for conductance. In this case, OF device acts as $\pm1$ defect. We have further discussed 2T2R structure and corresponding forming strategy that minimizes $\pm1$ defects. Having the optimized hardware that minimizes the critical initialization errors, we then evaluate the impact of the $\pm1$ and $0$ defects on the inference model accuracy. The numerical experiments show that the impact of $\pm1$ defect is more significant than that of the $0$ defects. Furthermore, we show that the trained model becomes resilient to the defects if we apply the same type of random defects on the weight matrices during the training. The defect-aware model shows smaller test error than the inference results from the baseline when $\pm1$ and $0$ defects are introduced. We also have discussed the logical value deviation of OF device ($\delta w_{OF}$) and FF device ($\delta w_{FF}$). Such deviation from the desired value occurs as both OF or FF device exhibits conductance which is deviated from the lowest/highest possible conductance value. We have discussed the reasonable range of such deviation and show that defect-aware model is also resilient to the errors occurred by $\delta w_{OF,FF}$. Lastly, we discuss a variation in the defect probability which may happen by the process variation or process drift. We have examined the impact of such variation by using an exemplary defect probability distribution. The result shows that including a known distribution of the defect probability during the training further improves the average inference performance as well as the standard deviation of the test error. 

One caveat of this study is that our analysis only includes the initialization failure scenario as the hardware non-ideality. In reality, the intricate interplay between different error sources may worsen the model accuracy and demands tighter parameter control than the scenario where individual components are separately considered \cite{2019Haensch}. 
To make a further statement on the realistic accuracy, future study needs to incorporate our results with the simulation framework that is capable of addressing other types of non-idealities. 

The yield improvement of the semiconductor industry is one of the most critical factors determining manufacturing cost. In this regard, understanding the major failure mechanism and co-optimize the hardware and software may provide an opportunity for sustainable incremental yield improvement on top of the device engineering and material innovations. The proposed optimization strategy provides a straightforward method to implement, and may open a new avenue to address similar problems for cross-bar array based on other types of devices (e.g. Phase change memory) or other types of neural network (e.g. recurrent network).

\section{Acknowledgments}
Y. Kim thanks to Theodorus E Standaert and Robert R. Robison for their managerial support. Y. Kim thanks to Wilfried Haensch and Geoffrey W. Burr for fruitful discussions. S. Kim acknowledges useful discussions from Tayfun Gokmen and managerial support from John Rozen. J. Choi thanks to Kailash Gopalakrishnan for his managerial support.

% The next two lines define the bibliography style to be used, and the bibliography file.
\bibliographystyle{ACM-Reference-Format}
\bibliography{inference_failure_ACM_v4_arxiv}


\begin{thebibliography}{37}

%%% the Web address.

\ifx \showCODEN    \undefined \def \showCODEN     #1{\unskip}     \fi
\ifx \showDOI      \undefined \def \showDOI       #1{#1}\fi
\ifx \showISBNx    \undefined \def \showISBNx     #1{\unskip}     \fi
\ifx \showISBNxiii \undefined \def \showISBNxiii  #1{\unskip}     \fi
\ifx \showISSN     \undefined \def \showISSN      #1{\unskip}     \fi
\ifx \showLCCN     \undefined \def \showLCCN      #1{\unskip}     \fi
\ifx \shownote     \undefined \def \shownote      #1{#1}          \fi
\ifx \showarticletitle \undefined \def \showarticletitle #1{#1}   \fi
\ifx \showURL      \undefined \def \showURL       {\relax}        \fi
% invisible to TeX
\providecommand\bibfield[2]{#2}
\providecommand\bibinfo[2]{#2}
\providecommand\natexlab[1]{#1}
\providecommand\showeprint[2][]{arXiv:#2}

\bibitem[\protect\citeauthoryear{Ambrogio, Narayanan, Tsai, Shelby, Boybat,
  Nolfo, Sidler, Giordano, Bodini, Farinha, et~al\mbox{.}}{Ambrogio
  et~al\mbox{.}}{2018}]%
        {ambrogio2018equivalent}
\bibfield{author}{\bibinfo{person}{Stefano Ambrogio}, \bibinfo{person}{Pritish
  Narayanan}, \bibinfo{person}{Hsinyu Tsai}, \bibinfo{person}{Robert~M Shelby},
  \bibinfo{person}{Irem Boybat}, \bibinfo{person}{Carmelo Nolfo},
  \bibinfo{person}{Severin Sidler}, \bibinfo{person}{Massimo Giordano},
  \bibinfo{person}{Martina Bodini}, \bibinfo{person}{Nathan~CP Farinha},
  {et~al\mbox{.}}} \bibinfo{year}{2018}\natexlab{}.
\newblock \showarticletitle{Equivalent-accuracy accelerated neural-network
  training using analogue memory}.
\newblock \bibinfo{journal}{\emph{Nature}} \bibinfo{volume}{558},
  \bibinfo{number}{7708} (\bibinfo{year}{2018}), \bibinfo{pages}{60}.
\newblock


\bibitem[\protect\citeauthoryear{Bersuker, Gilmer, and Veksler}{Bersuker
  et~al\mbox{.}}{2019}]%
        {bersuker2019metal}
\bibfield{author}{\bibinfo{person}{G Bersuker}, \bibinfo{person}{DC Gilmer},
  {and} \bibinfo{person}{D Veksler}.} \bibinfo{year}{2019}\natexlab{}.
\newblock \showarticletitle{Metal-oxide resistive random access memory (RRAM)
  technology: Material and operation details and ramifications}.
\newblock In \bibinfo{booktitle}{\emph{Advances in Non-Volatile Memory and
  Storage Technology}}. \bibinfo{publisher}{Elsevier},
  \bibinfo{pages}{35--102}.
\newblock


\bibitem[\protect\citeauthoryear{Bocquet, Hirztlin, Klein, Nowak, Vianello,
  Portal, and Querlioz}{Bocquet et~al\mbox{.}}{2018}]%
        {bocquet2018inmemory}
\bibfield{author}{\bibinfo{person}{M. Bocquet}, \bibinfo{person}{T. Hirztlin},
  \bibinfo{person}{J.~. Klein}, \bibinfo{person}{E. Nowak}, \bibinfo{person}{E.
  Vianello}, \bibinfo{person}{J.~. Portal}, {and} \bibinfo{person}{D.
  Querlioz}.} \bibinfo{year}{2018}\natexlab{}.
\newblock \showarticletitle{In-Memory and Error-Immune Differential ReRAM
  Implementation of Binarized Deep Neural Networks}. In
  \bibinfo{booktitle}{\emph{2018 IEEE International Electron Devices Meeting
  (IEDM)}}. \bibinfo{pages}{20.6.1--20.6.4}.
\newblock
\showISSN{2156-017X}
\urldef\tempurl%
\url{https://doi.org/10.1109/IEDM.2018.8614639}
\showDOI{\tempurl}


\bibitem[\protect\citeauthoryear{Chen}{Chen}{2013}]%
        {chen2013area}
\bibfield{author}{\bibinfo{person}{An Chen}.} \bibinfo{year}{2013}\natexlab{}.
\newblock \showarticletitle{Area and thickness scaling of forming voltage of
  resistive switching memories}.
\newblock \bibinfo{journal}{\emph{IEEE Electron Device Letters}}
  \bibinfo{volume}{35}, \bibinfo{number}{1} (\bibinfo{year}{2013}),
  \bibinfo{pages}{57--59}.
\newblock


\bibitem[\protect\citeauthoryear{Chen, Shih, Wu, Lin, Chiu, Sheu, and
  Chen}{Chen et~al\mbox{.}}{2014}]%
        {chen2014rram}
\bibfield{author}{\bibinfo{person}{Ching-Yi Chen}, \bibinfo{person}{Hsiu-Chuan
  Shih}, \bibinfo{person}{Cheng-Wen Wu}, \bibinfo{person}{Chih-He Lin},
  \bibinfo{person}{Pi-Feng Chiu}, \bibinfo{person}{Shyh-Shyuan Sheu}, {and}
  \bibinfo{person}{Frederick~T Chen}.} \bibinfo{year}{2014}\natexlab{}.
\newblock \showarticletitle{RRAM defect modeling and failure analysis based on
  march test and a novel squeeze-search scheme}.
\newblock \bibinfo{journal}{\emph{IEEE Trans. Comput.}} \bibinfo{volume}{64},
  \bibinfo{number}{1} (\bibinfo{year}{2014}), \bibinfo{pages}{180--190}.
\newblock


\bibitem[\protect\citeauthoryear{Chi, Li, Xu, Zhang, Zhao, Liu, Wang, and
  Xie}{Chi et~al\mbox{.}}{2016}]%
        {chi2016prime}
\bibfield{author}{\bibinfo{person}{Ping Chi}, \bibinfo{person}{Shuangchen Li},
  \bibinfo{person}{Cong Xu}, \bibinfo{person}{Tao Zhang},
  \bibinfo{person}{Jishen Zhao}, \bibinfo{person}{Yongpan Liu},
  \bibinfo{person}{Yu Wang}, {and} \bibinfo{person}{Yuan Xie}.}
  \bibinfo{year}{2016}\natexlab{}.
\newblock \showarticletitle{Prime: A novel processing-in-memory architecture
  for neural network computation in reram-based main memory}. In
  \bibinfo{booktitle}{\emph{ACM SIGARCH Computer Architecture News}},
  Vol.~\bibinfo{volume}{44}. IEEE Press, \bibinfo{pages}{27--39}.
\newblock


\bibitem[\protect\citeauthoryear{Choi, Wang, Venkataramani, Chuang, Srinivasan,
  and Gopalakrishnan}{Choi et~al\mbox{.}}{2018}]%
        {choi2018pact}
\bibfield{author}{\bibinfo{person}{Jungwook Choi}, \bibinfo{person}{Zhuo Wang},
  \bibinfo{person}{Swagath Venkataramani}, \bibinfo{person}{Pierce I-Jen
  Chuang}, \bibinfo{person}{Vijayalakshmi Srinivasan}, {and}
  \bibinfo{person}{Kailash Gopalakrishnan}.} \bibinfo{year}{2018}\natexlab{}.
\newblock \showarticletitle{PACT: Parameterized Clipping Activation for
  Quantized Neural Networks}.
\newblock \bibinfo{journal}{\emph{arXiv preprint arXiv:1805.06085}}
  (\bibinfo{year}{2018}).
\newblock


\bibitem[\protect\citeauthoryear{Courbariaux, Hubara, Soudry, El-Yaniv, and
  Bengio}{Courbariaux et~al\mbox{.}}{2016}]%
        {courbariaux2016binarized}
\bibfield{author}{\bibinfo{person}{Matthieu Courbariaux}, \bibinfo{person}{Itay
  Hubara}, \bibinfo{person}{Daniel Soudry}, \bibinfo{person}{Ran El-Yaniv},
  {and} \bibinfo{person}{Yoshua Bengio}.} \bibinfo{year}{2016}\natexlab{}.
\newblock \showarticletitle{Binarized neural networks: Training deep neural
  networks with weights and activations constrained to+ 1 or-1}.
\newblock \bibinfo{journal}{\emph{arXiv preprint arXiv:1602.02830}}
  (\bibinfo{year}{2016}).
\newblock


\bibitem[\protect\citeauthoryear{Gokmen, Onen, and Haensch}{Gokmen
  et~al\mbox{.}}{2017}]%
        {gokmen2018trainingConv}
\bibfield{author}{\bibinfo{person}{Tayfun Gokmen}, \bibinfo{person}{Murat
  Onen}, {and} \bibinfo{person}{Wilfried Haensch}.}
  \bibinfo{year}{2017}\natexlab{}.
\newblock \showarticletitle{Training Deep Convolutional Neural Networks with
  Resistive Cross-Point Devices}.
\newblock \bibinfo{journal}{\emph{Frontiers in Neuroscience}}
  \bibinfo{volume}{11} (\bibinfo{year}{2017}), \bibinfo{pages}{538}.
\newblock
\showISSN{1662-453X}
\urldef\tempurl%
\url{https://doi.org/10.3389/fnins.2017.00538}
\showDOI{\tempurl}


\bibitem[\protect\citeauthoryear{Gokmen, Rasch, and Haensch}{Gokmen
  et~al\mbox{.}}{2018}]%
        {gokmen2018trainingLSTM}
\bibfield{author}{\bibinfo{person}{Tayfun Gokmen}, \bibinfo{person}{Malte
  Rasch}, {and} \bibinfo{person}{Wilfried Haensch}.}
  \bibinfo{year}{2018}\natexlab{}.
\newblock \showarticletitle{Training LSTM Networks with Resistive Cross-Point
  Devices}.
\newblock \bibinfo{journal}{\emph{arXiv preprint arXiv:1806.00166}}
  (\bibinfo{year}{2018}).
\newblock


\bibitem[\protect\citeauthoryear{Gokmen and Vlasov}{Gokmen and Vlasov}{2016}]%
        {gokmen2016acceleration}
\bibfield{author}{\bibinfo{person}{Tayfun Gokmen} {and} \bibinfo{person}{Yurii
  Vlasov}.} \bibinfo{year}{2016}\natexlab{}.
\newblock \showarticletitle{Acceleration of deep neural network training with
  resistive cross-point devices: design considerations}.
\newblock \bibinfo{journal}{\emph{Frontiers in neuroscience}}
  \bibinfo{volume}{10} (\bibinfo{year}{2016}), \bibinfo{pages}{333}.
\newblock


\bibitem[\protect\citeauthoryear{Grossi, Zambelli, Olivo, Miranda, Stikanov,
  Walczyk, and Wenger}{Grossi et~al\mbox{.}}{2016}]%
        {grossi2016electrical}
\bibfield{author}{\bibinfo{person}{Alessandro Grossi},
  \bibinfo{person}{Cristian Zambelli}, \bibinfo{person}{Piero Olivo},
  \bibinfo{person}{Enrique Miranda}, \bibinfo{person}{Valeriy Stikanov},
  \bibinfo{person}{Christian Walczyk}, {and} \bibinfo{person}{Christian
  Wenger}.} \bibinfo{year}{2016}\natexlab{}.
\newblock \showarticletitle{Electrical characterization and modeling of
  pulse-based forming techniques in RRAM arrays}.
\newblock \bibinfo{journal}{\emph{Solid-State Electronics}}
  \bibinfo{volume}{115} (\bibinfo{year}{2016}), \bibinfo{pages}{17--25}.
\newblock


\bibitem[\protect\citeauthoryear{{Guo}, {Bayat}, {Bavandpour}, {Klachko},
  {Mahmoodi}, {Prezioso}, {Likharev}, and {Strukov}}{{Guo}
  et~al\mbox{.}}{2017}]%
        {Guo2017}
\bibfield{author}{\bibinfo{person}{X. {Guo}}, \bibinfo{person}{F.~M. {Bayat}},
  \bibinfo{person}{M. {Bavandpour}}, \bibinfo{person}{M. {Klachko}},
  \bibinfo{person}{M.~R. {Mahmoodi}}, \bibinfo{person}{M. {Prezioso}},
  \bibinfo{person}{K.~K. {Likharev}}, {and} \bibinfo{person}{D.~B. {Strukov}}.}
  \bibinfo{year}{2017}\natexlab{}.
\newblock \showarticletitle{Fast, energy-efficient, robust, and reproducible
  mixed-signal neuromorphic classifier based on embedded NOR flash memory
  technology}. In \bibinfo{booktitle}{\emph{2017 IEEE International Electron
  Devices Meeting (IEDM)}}. \bibinfo{pages}{6.5.1--6.5.4}.
\newblock
\showISSN{2156-017X}
\urldef\tempurl%
\url{https://doi.org/10.1109/IEDM.2017.8268341}
\showDOI{\tempurl}


\bibitem[\protect\citeauthoryear{{Haensch}, {Gokmen}, and {Puri}}{{Haensch}
  et~al\mbox{.}}{2019}]%
        {2019Haensch}
\bibfield{author}{\bibinfo{person}{W. {Haensch}}, \bibinfo{person}{T.
  {Gokmen}}, {and} \bibinfo{person}{R. {Puri}}.}
  \bibinfo{year}{2019}\natexlab{}.
\newblock \showarticletitle{The Next Generation of Deep Learning Hardware:
  Analog Computing}.
\newblock \bibinfo{journal}{\emph{Proc. IEEE}} \bibinfo{volume}{107},
  \bibinfo{number}{1} (\bibinfo{date}{Jan} \bibinfo{year}{2019}),
  \bibinfo{pages}{108--122}.
\newblock
\showISSN{0018-9219}
\urldef\tempurl%
\url{https://doi.org/10.1109/JPROC.2018.2871057}
\showDOI{\tempurl}


\bibitem[\protect\citeauthoryear{He, Zhang, Ren, and Sun}{He
  et~al\mbox{.}}{2016a}]%
        {he2016deep}
\bibfield{author}{\bibinfo{person}{Kaiming He}, \bibinfo{person}{Xiangyu
  Zhang}, \bibinfo{person}{Shaoqing Ren}, {and} \bibinfo{person}{Jian Sun}.}
  \bibinfo{year}{2016}\natexlab{a}.
\newblock \showarticletitle{Deep residual learning for image recognition}. In
  \bibinfo{booktitle}{\emph{Proceedings of the IEEE conference on computer
  vision and pattern recognition}}. \bibinfo{pages}{770--778}.
\newblock


\bibitem[\protect\citeauthoryear{He, Zhang, Ren, and Sun}{He
  et~al\mbox{.}}{2016b}]%
        {he2016identity}
\bibfield{author}{\bibinfo{person}{Kaiming He}, \bibinfo{person}{Xiangyu
  Zhang}, \bibinfo{person}{Shaoqing Ren}, {and} \bibinfo{person}{Jian Sun}.}
  \bibinfo{year}{2016}\natexlab{b}.
\newblock \showarticletitle{Identity mappings in deep residual networks}. In
  \bibinfo{booktitle}{\emph{European conference on computer vision}}. Springer,
  \bibinfo{pages}{630--645}.
\newblock


\bibitem[\protect\citeauthoryear{Jain, Ankit, Chakraborty, Gokmen, Rasch,
  Haensch, Roy, and Raghunathan}{Jain et~al\mbox{.}}{2019}]%
        {Jain2019NeuralNA}
\bibfield{author}{\bibinfo{person}{Shubham Jain}, \bibinfo{person}{Aayush
  Ankit}, \bibinfo{person}{Indranil Chakraborty}, \bibinfo{person}{Tayfun
  Gokmen}, \bibinfo{person}{Malte~J. Rasch}, \bibinfo{person}{Wilfried
  Haensch}, \bibinfo{person}{Kairshik Roy}, {and} \bibinfo{person}{Anand
  Raghunathan}.} \bibinfo{year}{2019}\natexlab{}.
\newblock \showarticletitle{Neural network accelerator design with resistive
  crossbars: Opportunities and challenges}.
\newblock \bibinfo{journal}{\emph{IBM Journal of Research and Development}}
  \bibinfo{volume}{63} (\bibinfo{year}{2019}), \bibinfo{pages}{10:1--10:13}.
\newblock


\bibitem[\protect\citeauthoryear{Kim, McIntyre, On~Chui, Saraswat, and
  Stemmer}{Kim et~al\mbox{.}}{2004}]%
        {kim2004engineering}
\bibfield{author}{\bibinfo{person}{Hyoungsub Kim}, \bibinfo{person}{Paul~C
  McIntyre}, \bibinfo{person}{Chi On~Chui}, \bibinfo{person}{Krishna~C
  Saraswat}, {and} \bibinfo{person}{Susanne Stemmer}.}
  \bibinfo{year}{2004}\natexlab{}.
\newblock \showarticletitle{Engineering chemically abrupt high-k metal oxide/
  silicon interfaces using an oxygen-gettering metal overlayer}.
\newblock \bibinfo{journal}{\emph{Journal of Applied Physics}}
  \bibinfo{volume}{96}, \bibinfo{number}{6} (\bibinfo{year}{2004}),
  \bibinfo{pages}{3467--3472}.
\newblock


\bibitem[\protect\citeauthoryear{Kinoshita, Tsunoda, Sato, Noshiro, Yagaki,
  Aoki, and Sugiyama}{Kinoshita et~al\mbox{.}}{2008}]%
        {kinoshita2008reduction}
\bibfield{author}{\bibinfo{person}{K Kinoshita}, \bibinfo{person}{K Tsunoda},
  \bibinfo{person}{Y Sato}, \bibinfo{person}{H Noshiro}, \bibinfo{person}{S
  Yagaki}, \bibinfo{person}{M Aoki}, {and} \bibinfo{person}{Y Sugiyama}.}
  \bibinfo{year}{2008}\natexlab{}.
\newblock \showarticletitle{Reduction in the reset current in a resistive
  random access memory consisting of Ni O x brought about by reducing a
  parasitic capacitance}.
\newblock \bibinfo{journal}{\emph{Applied Physics Letters}}
  \bibinfo{volume}{93}, \bibinfo{number}{3} (\bibinfo{year}{2008}),
  \bibinfo{pages}{033506}.
\newblock


\bibitem[\protect\citeauthoryear{Li, Hu, Li, Jiang, Ge, Montgomery, Zhang,
  Song, D{\'a}vila, Graves, et~al\mbox{.}}{Li et~al\mbox{.}}{2018}]%
        {li2018analogue}
\bibfield{author}{\bibinfo{person}{Can Li}, \bibinfo{person}{Miao Hu},
  \bibinfo{person}{Yunning Li}, \bibinfo{person}{Hao Jiang},
  \bibinfo{person}{Ning Ge}, \bibinfo{person}{Eric Montgomery},
  \bibinfo{person}{Jiaming Zhang}, \bibinfo{person}{Wenhao Song},
  \bibinfo{person}{Noraica D{\'a}vila}, \bibinfo{person}{Catherine~E Graves},
  {et~al\mbox{.}}} \bibinfo{year}{2018}\natexlab{}.
\newblock \showarticletitle{Analogue signal and image processing with large
  memristor crossbars}.
\newblock \bibinfo{journal}{\emph{Nature Electronics}} \bibinfo{volume}{1},
  \bibinfo{number}{1} (\bibinfo{year}{2018}), \bibinfo{pages}{52}.
\newblock


\bibitem[\protect\citeauthoryear{Liu, Lee, and Yu}{Liu et~al\mbox{.}}{2017}]%
        {liu2017analyzing}
\bibfield{author}{\bibinfo{person}{Rui Liu}, \bibinfo{person}{Heng-Yuan Lee},
  {and} \bibinfo{person}{Shimeng Yu}.} \bibinfo{year}{2017}\natexlab{}.
\newblock \showarticletitle{Analyzing inference robustness of ReRAM synaptic
  array in low-precision neural network}. In
  \bibinfo{booktitle}{\emph{Solid-State Device Research Conference (ESSDERC),
  2017 47th European}}. IEEE, \bibinfo{pages}{18--21}.
\newblock


\bibitem[\protect\citeauthoryear{Long, Na, and Mukhopadhyay}{Long
  et~al\mbox{.}}{2018}]%
        {long2018reram}
\bibfield{author}{\bibinfo{person}{Yun Long}, \bibinfo{person}{Taesik Na},
  {and} \bibinfo{person}{Saibal Mukhopadhyay}.}
  \bibinfo{year}{2018}\natexlab{}.
\newblock \showarticletitle{ReRAM-based processing-in-memory architecture for
  recurrent neural network acceleration}.
\newblock \bibinfo{journal}{\emph{IEEE Transactions on Very Large Scale
  Integration (VLSI) Systems}} \bibinfo{number}{99} (\bibinfo{year}{2018}),
  \bibinfo{pages}{1--14}.
\newblock


\bibitem[\protect\citeauthoryear{{Merrikh-Bayat}, {Guo}, {Klachko}, {Prezioso},
  {Likharev}, and {Strukov}}{{Merrikh-Bayat} et~al\mbox{.}}{2018}]%
        {Merrikh2018}
\bibfield{author}{\bibinfo{person}{F. {Merrikh-Bayat}}, \bibinfo{person}{X.
  {Guo}}, \bibinfo{person}{M. {Klachko}}, \bibinfo{person}{M. {Prezioso}},
  \bibinfo{person}{K.~K. {Likharev}}, {and} \bibinfo{person}{D.~B. {Strukov}}.}
  \bibinfo{year}{2018}\natexlab{}.
\newblock \showarticletitle{High-Performance Mixed-Signal Neurocomputing With
  Nanoscale Floating-Gate Memory Cell Arrays}.
\newblock \bibinfo{journal}{\emph{IEEE Transactions on Neural Networks and
  Learning Systems}} \bibinfo{volume}{29}, \bibinfo{number}{10}
  (\bibinfo{date}{Oct} \bibinfo{year}{2018}), \bibinfo{pages}{4782--4790}.
\newblock
\showISSN{2162-237X}
\urldef\tempurl%
\url{https://doi.org/10.1109/TNNLS.2017.2778940}
\showDOI{\tempurl}


\bibitem[\protect\citeauthoryear{Pan, Gao, Chen, Song, and Zeng}{Pan
  et~al\mbox{.}}{2014}]%
        {pan2014recent}
\bibfield{author}{\bibinfo{person}{Feng Pan}, \bibinfo{person}{Shuang Gao},
  \bibinfo{person}{Chao Chen}, \bibinfo{person}{C Song}, {and}
  \bibinfo{person}{F Zeng}.} \bibinfo{year}{2014}\natexlab{}.
\newblock \showarticletitle{Recent progress in resistive random access
  memories: materials, switching mechanisms, and performance}.
\newblock \bibinfo{journal}{\emph{Materials Science and Engineering: R:
  Reports}}  \bibinfo{volume}{83} (\bibinfo{year}{2014}),
  \bibinfo{pages}{1--59}.
\newblock


\bibitem[\protect\citeauthoryear{Rastegari, Ordonez, Redmon, and
  Farhadi}{Rastegari et~al\mbox{.}}{2016}]%
        {rastegari2016xnor}
\bibfield{author}{\bibinfo{person}{Mohammad Rastegari},
  \bibinfo{person}{Vicente Ordonez}, \bibinfo{person}{Joseph Redmon}, {and}
  \bibinfo{person}{Ali Farhadi}.} \bibinfo{year}{2016}\natexlab{}.
\newblock \showarticletitle{Xnor-net: Imagenet classification using binary
  convolutional neural networks}. In \bibinfo{booktitle}{\emph{European
  Conference on Computer Vision}}. Springer, \bibinfo{pages}{525--542}.
\newblock


\bibitem[\protect\citeauthoryear{Romero, Ambrogio, Giordano, Cristiano, Bodini,
  Narayanan, Tsai, Shelby, and Burr}{Romero et~al\mbox{.}}{2019}]%
        {romero2019training}
\bibfield{author}{\bibinfo{person}{Louis~P Romero}, \bibinfo{person}{Stefano
  Ambrogio}, \bibinfo{person}{Massimo Giordano}, \bibinfo{person}{Giorgio
  Cristiano}, \bibinfo{person}{Martina Bodini}, \bibinfo{person}{Pritish
  Narayanan}, \bibinfo{person}{Hsinyu Tsai}, \bibinfo{person}{Robert~M Shelby},
  {and} \bibinfo{person}{Geoffrey~W Burr}.} \bibinfo{year}{2019}\natexlab{}.
\newblock \showarticletitle{Training fully connected networks with resistive
  memories: Impact of device failures}.
\newblock \bibinfo{journal}{\emph{Faraday Discussions}}  \bibinfo{volume}{213}
  (\bibinfo{year}{2019}), \bibinfo{pages}{371--391}.
\newblock


\bibitem[\protect\citeauthoryear{Shafiee, Nag, Muralimanohar, Balasubramonian,
  Strachan, Hu, Williams, and Srikumar}{Shafiee et~al\mbox{.}}{2016}]%
        {shafiee2016isaac}
\bibfield{author}{\bibinfo{person}{Ali Shafiee}, \bibinfo{person}{Anirban Nag},
  \bibinfo{person}{Naveen Muralimanohar}, \bibinfo{person}{Rajeev
  Balasubramonian}, \bibinfo{person}{John~Paul Strachan}, \bibinfo{person}{Miao
  Hu}, \bibinfo{person}{R~Stanley Williams}, {and} \bibinfo{person}{Vivek
  Srikumar}.} \bibinfo{year}{2016}\natexlab{}.
\newblock \showarticletitle{ISAAC: A convolutional neural network accelerator
  with in-situ analog arithmetic in crossbars}.
\newblock \bibinfo{journal}{\emph{ACM SIGARCH Computer Architecture News}}
  \bibinfo{volume}{44}, \bibinfo{number}{3} (\bibinfo{year}{2016}),
  \bibinfo{pages}{14--26}.
\newblock


\bibitem[\protect\citeauthoryear{Shih, Chen, Wu, Lin, and Sheu}{Shih
  et~al\mbox{.}}{2011}]%
        {shih2011training}
\bibfield{author}{\bibinfo{person}{Hsiu-Chuan Shih}, \bibinfo{person}{Ching-Yi
  Chen}, \bibinfo{person}{Cheng-Wen Wu}, \bibinfo{person}{Chih-He Lin}, {and}
  \bibinfo{person}{Shyh-Shyuan Sheu}.} \bibinfo{year}{2011}\natexlab{}.
\newblock \showarticletitle{Training-based forming process for RRAM yield
  improvement}. In \bibinfo{booktitle}{\emph{29th VLSI Test Symposium}}. IEEE,
  \bibinfo{pages}{146--151}.
\newblock


\bibitem[\protect\citeauthoryear{Song, Qian, Li, and Chen}{Song
  et~al\mbox{.}}{2017}]%
        {song2017pipelayer}
\bibfield{author}{\bibinfo{person}{Linghao Song}, \bibinfo{person}{Xuehai
  Qian}, \bibinfo{person}{Hai Li}, {and} \bibinfo{person}{Yiran Chen}.}
  \bibinfo{year}{2017}\natexlab{}.
\newblock \showarticletitle{Pipelayer: A pipelined reram-based accelerator for
  deep learning}. In \bibinfo{booktitle}{\emph{High Performance Computer
  Architecture (HPCA), 2017 IEEE International Symposium on}}. IEEE,
  \bibinfo{pages}{541--552}.
\newblock


\bibitem[\protect\citeauthoryear{Sun, Yin, Peng, Liu, Seo, and Yu}{Sun
  et~al\mbox{.}}{[n. d.]}]%
        {sun2018xnor}
\bibfield{author}{\bibinfo{person}{Xiaoyu Sun}, \bibinfo{person}{Shihui Yin},
  \bibinfo{person}{Xiaochen Peng}, \bibinfo{person}{Rui Liu},
  \bibinfo{person}{Jae-sun Seo}, {and} \bibinfo{person}{Shimeng Yu}.}
  \bibinfo{year}{[n. d.]}\natexlab{}.
\newblock \showarticletitle{XNOR-ReRAM: A scalable and parallel resistive
  synaptic architecture for binary neural networks}.
\newblock \bibinfo{journal}{\emph{algorithms}}  \bibinfo{volume}{2}
  (\bibinfo{year}{[n. d.]}), \bibinfo{pages}{3}.
\newblock


\bibitem[\protect\citeauthoryear{Thompson and Spanuth}{Thompson and
  Spanuth}{2018}]%
        {thompson2018decline}
\bibfield{author}{\bibinfo{person}{Neil Thompson} {and} \bibinfo{person}{Svenja
  Spanuth}.} \bibinfo{year}{2018}\natexlab{}.
\newblock \showarticletitle{The Decline of Computers As a General Purpose
  Technology: Why Deep Learning and the End of Moore’s Law are Fragmenting
  Computing}.
\newblock \bibinfo{journal}{\emph{Available at SSRN 3287769}}
  (\bibinfo{year}{2018}).
\newblock


\bibitem[\protect\citeauthoryear{Villarrubia, De~Paz, Chamoso, and De~la
  Prieta}{Villarrubia et~al\mbox{.}}{2018}]%
        {villarrubia2018artificial}
\bibfield{author}{\bibinfo{person}{Gabriel Villarrubia},
  \bibinfo{person}{Juan~F De~Paz}, \bibinfo{person}{Pablo Chamoso}, {and}
  \bibinfo{person}{Fernando De~la Prieta}.} \bibinfo{year}{2018}\natexlab{}.
\newblock \showarticletitle{Artificial neural networks used in optimization
  problems}.
\newblock \bibinfo{journal}{\emph{Neurocomputing}}  \bibinfo{volume}{272}
  (\bibinfo{year}{2018}), \bibinfo{pages}{10--16}.
\newblock


\bibitem[\protect\citeauthoryear{Xia, Huangfu, Tang, Yin, Chakrabarty, Xie,
  Wang, and Yang}{Xia et~al\mbox{.}}{2017}]%
        {xia2017stuck}
\bibfield{author}{\bibinfo{person}{Lixue Xia}, \bibinfo{person}{Wenqin
  Huangfu}, \bibinfo{person}{Tianqi Tang}, \bibinfo{person}{Xiling Yin},
  \bibinfo{person}{Krishnendu Chakrabarty}, \bibinfo{person}{Yuan Xie},
  \bibinfo{person}{Yu Wang}, {and} \bibinfo{person}{Huazhong Yang}.}
  \bibinfo{year}{2017}\natexlab{}.
\newblock \showarticletitle{Stuck-at fault tolerance in RRAM computing
  systems}.
\newblock \bibinfo{journal}{\emph{IEEE Journal on Emerging and Selected Topics
  in Circuits and Systems}} \bibinfo{volume}{8}, \bibinfo{number}{1}
  (\bibinfo{year}{2017}), \bibinfo{pages}{102--115}.
\newblock


\bibitem[\protect\citeauthoryear{Zhang, Pezeshki, Brakel, Zhang, Laurent,
  Bengio, and Courville}{Zhang et~al\mbox{.}}{2016}]%
        {zhang2016towards}
\bibfield{author}{\bibinfo{person}{Ying Zhang}, \bibinfo{person}{Mohammad
  Pezeshki}, \bibinfo{person}{Phil{\'e}mon Brakel}, \bibinfo{person}{Saizheng
  Zhang}, \bibinfo{person}{C{\'e}sar Laurent}, \bibinfo{person}{Yoshua Bengio},
  {and} \bibinfo{person}{Aaron Courville}.} \bibinfo{year}{2016}\natexlab{}.
\newblock \showarticletitle{Towards End-to-End Speech Recognition with Deep
  Convolutional Neural Networks}.
\newblock \bibinfo{journal}{\emph{Interspeech 2016}} (\bibinfo{year}{2016}),
  \bibinfo{pages}{410--414}.
\newblock


\bibitem[\protect\citeauthoryear{Zhou, Wu, Ni, Zhou, Wen, and Zou}{Zhou
  et~al\mbox{.}}{2016}]%
        {zhou2016dorefa}
\bibfield{author}{\bibinfo{person}{Shuchang Zhou}, \bibinfo{person}{Yuxin Wu},
  \bibinfo{person}{Zekun Ni}, \bibinfo{person}{Xinyu Zhou}, \bibinfo{person}{He
  Wen}, {and} \bibinfo{person}{Yuheng Zou}.} \bibinfo{year}{2016}\natexlab{}.
\newblock \showarticletitle{Dorefa-net: Training low bitwidth convolutional
  neural networks with low bitwidth gradients}.
\newblock \bibinfo{journal}{\emph{arXiv preprint arXiv:1606.06160}}
  (\bibinfo{year}{2016}).
\newblock


\bibitem[\protect\citeauthoryear{Zhou, Huang, Xiang, Shen, Zhao, Feng, Gao, Wu,
  Qian, Liu, Zhang, Liu, and Kang}{Zhou et~al\mbox{.}}{2018}]%
        {zhou2018anewhardware}
\bibfield{author}{\bibinfo{person}{Z. Zhou}, \bibinfo{person}{P. Huang},
  \bibinfo{person}{Y.~C. Xiang}, \bibinfo{person}{W.~S. Shen},
  \bibinfo{person}{Y.~D. Zhao}, \bibinfo{person}{Y.~L. Feng},
  \bibinfo{person}{B. Gao}, \bibinfo{person}{H.~Q. Wu}, \bibinfo{person}{H.
  Qian}, \bibinfo{person}{L.~F. Liu}, \bibinfo{person}{X. Zhang},
  \bibinfo{person}{X.~Y. Liu}, {and} \bibinfo{person}{J.~F. Kang}.}
  \bibinfo{year}{2018}\natexlab{}.
\newblock \showarticletitle{A new hardware implementation approach of BNNs
  based on nonlinear 2T2R synaptic cell}. In \bibinfo{booktitle}{\emph{2018
  IEEE International Electron Devices Meeting (IEDM)}}.
  \bibinfo{pages}{20.7.1--20.7.4}.
\newblock
\showISSN{2156-017X}
\urldef\tempurl%
\url{https://doi.org/10.1109/IEDM.2018.8614642}
\showDOI{\tempurl}


\bibitem[\protect\citeauthoryear{Zmora, Jacob, and Novik}{Zmora
  et~al\mbox{.}}{2018}]%
        {distiller2018}
\bibfield{author}{\bibinfo{person}{Neta Zmora}, \bibinfo{person}{Guy Jacob},
  {and} \bibinfo{person}{Gal Novik}.} \bibinfo{year}{2018}\natexlab{}.
\newblock \bibinfo{title}{Neural Network Distiller}.
\newblock
\newblock
\urldef\tempurl%
\url{https://doi.org/10.5281/zenodo.1297430}
\showDOI{\tempurl}


\end{thebibliography}
%\bibliography{reference}

\end{document}